\documentclass[twocolumn,showpacs,preprintnumbers,amsmath,amssymb]{revtex4}
\usepackage{amsfonts}
\usepackage{amssymb}
\usepackage{psfig}
\newcommand{\beq}{\begin{equation}}
\newcommand{\eeq}{\end{equation}}
\newcommand{\beqn}{\begin{eqnarray}}
\newcommand{\eeqn}{\end{eqnarray}}
\newcommand{\bea}[1]{\beq\begin{array}{#1}}
\newcommand{\eea}{\end{array}\eeq}

\newcommand{\sign}{\mathop{\rm sign}}

\newcommand{\tr}{\mathop{\rm Tr}}

\newcommand{\ket}[1]{|\,#1\,\rangle}
\newcommand{\bra}[1]{\langle\,#1\,|}
\newcommand{\braket}[2]{\langle\,#1\,|\,#2\,\rangle}

\newcommand{\diff}{\partial}

\newcommand{\cT}{{\cal T}}

\newcommand{\HP}[1]{{\mathrm{HP}^#1}}
\newcommand{\hp}{\mathrm{HP}}

\begin{document}

\preprint{ITEP-LAT/2005-24}

\title{SU(2) Gluodynamics and $\HP{1}$ $\sigma$-model Embedding:\\
       Scaling, Topology and Confinement}

\author{P.Yu.~Boyko}
  \email{boyko@itep.ru}
\author{F.V.~Gubarev}
  \email{gubarev@itep.ru}
\author{S.M.~Morozov}
  \email{smoroz@itep.ru}

\affiliation{Institute of Theoretical and  Experimental Physics,
              B.Cheremushkinskaya 25, Moscow, 117218, Russia}

\begin{abstract}
We investigate recently proposed $\HP{1}$ $\sigma$-model embedding method
aimed to study the topology of SU(2) gauge fields.
The $\HP{1}$ based topological charge is shown to be fairly compatible with various known definitions.
We study the corresponding topological susceptibility and estimate its value in the continuum limit.
The geometrical clarity of $\HP{1}$ approach allows to investigate non-perturbative aspects
of SU(2) gauge theory on qualitatively new level. In particular, we obtain numerically precise
estimation of gluon condensate and its leading quadratic correction. Furthermore, we present
clear evidences that the string tension is to be associated with global (percolating)
regions of sign-coherent topological charge. As a byproduct of our analysis we estimate
the continuum value of quenched chiral condensate and the dimensionality of regions, which
localize the lowest eigenmodes of overlap Dirac operator.
\end{abstract}


\maketitle
\section{Introduction}
\label{section:intro}
The topological aspects of gauge theories had always been the prime topic for the lattice
community. However, the real breakthrough here is due to the construction of
chirally symmetric overlap Dirac operator on the lattice~\cite{Neuberger:1997fp},
which via Atiyah-Singer index theorem~\cite{Atiyah:1971rm} applied in lattice settings~\cite{overlap}
allowed to investigate the topology of equilibrium vacuum fields.
Since then a lot of remarkable results were obtained, among which 
the discovery of global (percolating) regions of sign-coherent topological
charge~\cite{Horvath-structures,Gubarev:2005rs}
is worth to be mentioned. Note that the overlap-based approach to the gauge fields topology
is rather involved technically, however, the intricacy of the method is not the only problem here.
What is more important for us is the factual absence of geometrically clean microscopic
picture behind the topological fluctuations in the overlap approach, which
is in a sharp contrast with what is known from seminal papers~\cite{Atiyah:1978ri,Drinfeld:1978xr}
about the topology of Yang-Mills fields (see also \cite{HPN-various}).

Recently the alternative definition of the topological charge~\cite{Gubarev:2005rs}
has been put forward, which is based essentially on the same topological ideas as were
used in~\cite{Atiyah:1978ri,Drinfeld:1978xr}.
The essence of the method consists in finding the nearest (in the configuration space)
to given SU(2) gauge background configuration of scalar fields representing non-linear
$\sigma$-model with target space being the quaternionic projective space $\HP{n}$
(see Ref.~\cite{Gubarev:2005rs} for missing details).
The relevance of $\HP{n}$ spaces is by no means accidental, the close connection between
$SU(2)$ Yang-Mills theory and $\HP{n}$ $\sigma$-models is known for
edges~\cite{Atiyah:1978ri,Drinfeld:1978xr,HPN-various,Dubrovin}.
Generically the idea is to realize the original $SU(2)$ gauge holonomy assigned
to every lattice link as the non-Abelian geometrical phase factor associated 
with motion of scalar particle possessing non-trivial internal symmetry space.
However, except for classical gauge backgrounds it is not possible to achieve this exactly,
hence the appearance of the term 'nearest configuration' above. 
Unfortunately, this is the weakest aspect of $\HP{n}$ $\sigma$-model approach since
the notion of nearest configuration could not, in fact, be conceived rigorously
in the case of quantum fields and we have to rely mostly on the results of numerical simulations.
However, we note that this is not a fatal restriction of the method, essentially the same problem
exists, for instance, in overlap-based approach, where it was by no means evident {\it a priori}
that chirally symmetric Dirac operator applied to thermalized fields is indeed local.  Note that for equilibrium
configurations the dependence upon the $\sigma$-model rank $n$ was found~\cite{Gubarev:2005rs}
to be trivial~\footnote{
This is not an unexpected result, justification could be found already in~\cite{Atiyah:1978ri,Drinfeld:1978xr}
(see Ref.~\cite{Gubarev:2005rs} for detailed discussion).
} and for this reason we consider only the case $n=1$ below.

The crucial advantage of $\HP{1}$ approach is its clean geometrical meaning, which not only allows
to capture the gauge fields topology, but also permits the microscopic investigation 
of the topological fluctuations -- the luxury which was not available so far.
The prime purpose of the present publication is to put this assertion on numerically
solid grounds. Note that some remarkable results we obtained already in Ref.~\cite{Gubarev:2005rs}
and we by no means casting doubts on them. 
However, as far as the phenomenologically relevant quantities are concerned, the numerics should
evidently be improved and we do this in present paper.
In particular, after short theoretical introduction and description
of the numerical algorithms (section~\ref{section:theory})
we present in section~\ref{section:numerics}
statistically convincing comparison of $\HP{1}$ and overlap-based topological charges
both measured on the same thermalized vacuum configurations. Furthermore, in section~\ref{section:chi}
the scaling of $\HP{1}$ based topological susceptibility with lattice spacing and volume
is discussed in details and its rather unambiguous extrapolation to the continuum limit is presented.
The outcome of this calculations makes us confident that $\HP{1}$ topological charge
is fairly compatible with overlap-based definition and is insensitive to lattice dislocations.
Moreover, keeping in mind the technical complexity and geometrical obscurity of the overlap construction,
we dare to claim that the $\HP{1}$ $\sigma$-model embedding method is indeed advantageous
both geometrically and computationally.

The geometrical explicitness of $\HP{1}$ approach is crucial for the remaining part of our paper.
Namely it allows us to investigate the topology related content of the original gauge fields
in explicit manner by introducing the so called $\HP{1}$ projection to be discussed in details
in section~\ref{section:HP}.
Note that the term 'projection' here has nothing in common with its nowadays conventional
meaning, nevertheless we use it mostly due to the historical reasons.
The results we obtain this way are indeed remarkable, however, before mentioning them we would
like note the following.
In this paper we intentionally leave aside any serious attempts to interpret
the experimental data. In fact, the theoretical interpretation could indeed be given,
moreover, it overlaps partially with moderns trends in the
literature~\cite{Gubarev:2005jm,Polikarpov:2005ey,A2,A2-review,viz,viz2}.
However, consideration of the theoretical implications would leads us to quite lengthy
discussions and for that reason we decided to focus exclusively on the experimental side
of the problem. Our theoretical interpretation of the results presented here
is to be published elsewhere~\cite{future}.

Section~\ref{section:HP} is devoted to the investigation of the dynamics of $\HP{1}$ projected fields obtained
from vacuum configurations at various lattice spacings. We start in section~\ref{section:HP-plaq}
from the consideration of the curvature of projected fields, which turns out to be extremely small.
Furthermore, its spacing dependence does not contain any sign whatsoever of the perturbation
theory contribution. Instead, the curvature vanishes in the limit $a\to 0$, the leading spacing
corrections being quartic and quadratic in $a$. In turn, these leading terms are to be interpreted
as the gluon condensate and quadratic power correction to it, for both of which we obtain
numerically precise estimations and compare them with available literature.
Then in section~\ref{section:HP-confinement}
the confining properties of $\HP{1}$ projected fields are investigated from various viewpoints.
The conclusion is that these fields are indeed confining, moreover, the corresponding string
tension accounts for the full $SU(2)$ string tension in the continuum limit.
The geometrical clarity of the $\HP{1}$ approach allows us to investigate in section~\ref{section:HP-confinement-lumps}
the microscopic origin of the projected string tension. We present clear
evidences that the string tension is to be associated with global (percolating) regions of
sign-coherent topological charge mentioned above.
Finally, to conclude the study of $\HP{1}$ projected fields dynamics we consider
in section~\ref{section:HP-fermions}
the low lying spectrum of overlap Dirac operator in the projected background and show that
both the spectral density and the localization properties of low lying modes are essentially
the same as they are on the original fields. The quenched chiral condensate and the dimensionality
of localization regions as they are seen within $\HP{1}$ projection are estimated.

The results of section~\ref{section:HP}
could only be convincing provided that we make sure that the gauge fields components
not surviving $\HP{1}$ projection do not contribute to neither of non-perturbative observables
discussed above. The problem is non-trivial due to the explicit gauge covariance of $\HP{1}$
projection, however we believe that it is addressed adequately in section~\ref{section:coset},
where we show that this part of the original gauge fields could be defined unambiguously.
Then it is the matter of straightforward calculation to show that the correspondingly constructed
configurations are:

{\it i)} topologically trivial, in a sense that the topological charge in both $\HP{1}$ and overlap
definitions vanishes identically on all configurations we have in our disposal;

{\it ii)} deconfining, the string tension vanishes while the heavy quark potential is still compatible
with Coulomb law;

{\it iii)} trivial from the fermionic viewpoint, the low lying spectrum of the Dirac operator
disappears completely, we were unable to find even a single low lying eigenmode on all our configurations.

These findings are the most stringent evidences that what had been left aside by $\HP{1}$ projection
corresponds to pure perturbation theory. At the same time, they show that the results and estimations of
various non-perturbative observables, which were obtained in section~\ref{section:HP},
are indeed reliable. Keeping in mind that the actual numbers are in complete agreement
with the existing literature, we dare to claim that the $\HP{1}$ projection is, in fact, the 
unique method allowing to achieve numerically most accurate results.

Finally, let us mention that the approach developed in sections~\ref{section:HP}, \ref{section:coset}
is parallel in spirit to what is known in the literature as 'vortex removal' \cite{vortex-removal} 
and 'monopole removal' \cite{Bornyakov:2005wp}
procedures. However, we stress that the $\HP{1}$ projection is by no means similar to them
both physically and technically.

\section{Theoretical Background and Numerical Algorithms}
\label{section:theory}

In this section we briefly outline the recent approach~\cite{Gubarev:2005rs}
aimed to investigate $SU(2)$ gauge fields topology. This material is included to introduce
the necessary notations and  to make the paper self-contained (for details and further
references see~\cite{Gubarev:2005rs}). In particular, in this section we precisely describe
the numerical methods utilized in our paper. Note that the present numerical implementation
is slightly different from that of~\cite{Gubarev:2005rs} although we checked that the difference
is completely inessential physics-wise.

Quaternionic projective space  $\HP{1}$ can be viewed as the factor space
$\HP{1} = S^4 = S^7 / S^3$ (second Hopf fibering) and its lowest non-trivial homotopic group is
$\pi_4(\HP{1}) = \pi_3(S^3) = Z$.
The simplest explicit parametrization of $\HP{1}$ is provided by normalized quaternionic vectors
\beq
\ket{q} = [q_0, q_1]^T\,, \quad  q_i\in \mathrm{H}\,,\quad
\braket{q}{q} = \bar{q}_i q_i = 1 \in \mathrm{H}\,,
\eeq
where  $[...]^T$ indicates transposition, $\mathrm{H}$ is the field of real quaternions,
$q_i = q_i^{\alpha} e_\alpha \in \mathrm{H}$, $\alpha = 0,...,3$ and $e_\alpha$ are the quaternionic units
\beq
e_0 = 1\,,\quad
e_i e_j = -\delta_{ij} - \varepsilon_{ijk} e_k\,, \,\,\,\,i,j,k = 1,2,3
\eeq
with conjugation defined as usual
\beq
\bar{q}_i = q_i^{\alpha} \bar{e}_\alpha\,, \qquad
\bar{e}_0 = e_0\,,\,\,\,\, \bar{e}_i = -e_i\,,\,\,\,\, i=1,2,3\,.
\eeq
The states $\ket{q}$ describe $7$-dimensional sphere, $\ket{q} \in S^7$,
while the $\HP{1}$ space is the set of equivalence classes of $\ket{q}$ with respect
to the right multiplication by unit quaternions (elements of SU(2) group)
\beq
\label{theory:gauge}
\ket{q} \sim \ket{q} \, \upsilon\,,\qquad |\upsilon|^2 \equiv \bar{\upsilon} \upsilon = 1\,,
\quad \upsilon\in\mathrm{H}\,.
\eeq
The gauge invariance (\ref{theory:gauge}) allows to introduce almost everywhere
inhomogeneous coordinate $\omega = q_1 q^{-1}_0 = y^\alpha e_\alpha \in \mathrm{H}$
on $\HP{1}$ space
\beq
\label{theory:inhomo}
\ket{q} = [q_0, q_1]^T ~ \sim ~ \frac{1}{\sqrt{1+|\omega|^2}}\,[ 1, \omega]^T\,.
\eeq
The alternative parametrization is provided
by 5-dimensional unit vector $n^A$, $A=0,...,4$ defined by
\beq
\label{theory:n}
n^A = \bra{q} \gamma^A \ket{q}\,,
\eeq
where $\gamma^A$ are five Euclidean Dirac matrices $\{\gamma^\alpha,\gamma^5\}$
viewed as $2\times 2$ quaternionic ones.
One could show that (\ref{theory:n}) is equivalent to the standard stereographic projection
which relates $n^A$ with inhomogeneous coordinate $\omega = y^\alpha e_\alpha$ above
\beq
n^\alpha = \frac{2 y^\alpha}{ 1 + y^2}\,, \,\,\,\,\alpha=0,...,3\,,
\qquad
n^4 = \frac{1 - y^2}{1 + y^2}\,.
\eeq
It is clear that non-linear $\sigma$-model with target space $\HP{1}$ would be non-trivial
provided that the base space is taken to be 4-sphere $S^4$. The corresponding topological
charge
\beqn
\label{theory:Q-n}
Q = \frac{1}{(8\pi)^2} \int d^4 x \,\, \varepsilon^{\mu\nu\lambda\rho}
    \, \varepsilon_{ABCDE} \cdot ~~~~~~~~~~~~~~~~~~~ \\
~~~~~~~~~~~~~~~~~~~ \cdot n^A \diff_\mu n^B \diff_\nu n^C \diff_\lambda n^D\diff_\rho n^E \nonumber
\eeqn
is geometrically the sum of oriented infinitesimal volumes in the image of $n^A(x) : S^4 \to \HP{1}$. 

As usual it is convenient to introduce auxiliary SU(2) gauge fields 
\beq
\label{theory:gauge-fields}
A_\mu = -A^\dagger_\mu = - \bra{q} \diff_\mu \ket{q}\,,
\eeq
transforming as $A_\mu \to \bar{\upsilon} A_\mu \upsilon - \bar{\upsilon}\diff_\mu\upsilon$ under
(\ref{theory:gauge}). Then the topological charge (\ref{theory:Q-n}) is expressible solely in terms of $A_\mu$
\beqn
\label{theory:Q-A}
Q = \frac{1}{32\pi^2} \int d^4 x \,\, \varepsilon^{\mu\nu\lambda\rho}\,\, \mathrm{Tr} \, F_{\mu\nu} F_{\lambda\rho}\,, \\
F_{\mu\nu} = \diff_\mu A_\nu - \diff_\nu A_\mu + [A_\mu, A_\nu]\,, \nonumber
\eeqn
being essentially equivalent to the familiar topological charge of the gauge fields (\ref{theory:gauge-fields}).
Note that Eq.~(\ref{theory:Q-A}) is by no means accidental, 
the deep connection between $\HP{n}$ $\sigma$-models and SU(2) Yang-Mills
theory is known for edges (see, e.g., Refs.~\cite{Atiyah:1978ri,Drinfeld:1978xr,HPN-various,Dubrovin}).
In particular, the geometry of gauge fields could best be analyzed
in the $\HP{n}$ $\sigma$-models context. In fact, all known instantonic solutions of Yang-Mills theory
could be considered as induced by the topological configurations of suitable $\HP{n}$ $\sigma$-models.

As is discussed in length in Ref.~\cite{Gubarev:2005rs}, 
it is natural to expect that the construction of nearest (in the configuration space) 
to the given gauge background $\HP{n}$ fields captures accurately the gauge fields topology leaving
aside the non-topological properties of the background. Note that for classical gauge fields the corresponding
$\HP{n}$ $\sigma$-model reproducing Eq.~(\ref{theory:gauge-fields}) is known to be unique~\cite{HPN-various}.
However, the mathematical rigour is lost in case of 'hot' vacuum configurations and we could only hope
to find the nearest in the sense of Eq.~(\ref{theory:gauge-fields}) $\sigma$-model fields.
The proposition of Ref.~\cite{Gubarev:2005rs} was to minimize the functional
\beq
F(A, q) ~=~ \int \left( A_\mu ~+~ \bra{q}\diff_\mu\ket{q} \right)^2
\eeq
for given $A_\mu$ with respect to all possible configurations of $\ket{q_x}$.
On the lattice this equation translates into
\beq
\label{theory:functional}
F(U, q) ~=~ 1 - \frac{1}{4 V} \sum\limits_{x,\mu} \mathrm{Sc}\left[ 
\frac{\braket{q_x}{q_{x+\mu}}}{|\braket{q_x}{q_{x+\mu}}|} \, U^\dagger_{x,\mu}\right]\,,
\eeq
where lattice gauge fields are denoted by $U_{x,\mu} \in SU(2)$, $V$ is the lattice volume
and quaternionic scalar part is $\mathrm{Sc}[q] = (q + \bar q)/2$.
The configuration $\ket{q_x}$ which provides the minimum to $F(U, q)$ is the best possible $\HP{n}$
fields approximation to the original gauge potentials
\beq
\label{theory:approx}
U_{x,\mu} ~\approx~ U^\hp_{x,\mu} ~\equiv~ \frac{\braket{q_x}{q_{x+\mu}}}{|\braket{q_x}{q_{x+\mu}}|}\,,
\eeq
while the minimum value of (\ref{theory:functional}) is a natural measure of the approximation quality.
Furthermore,  it was shown in Ref.~\cite{Gubarev:2005rs} that for equilibrium gauge configurations
the dependence on the $\sigma$-model rank $n$ is trivial since the quality of the approximation
(\ref{theory:approx}) does not depend on $n$. Therefore, it seems to be legitimate to consider
solely the case $n=1$ and only $\HP{1}$ $\sigma$-model embedding is investigated below.
Note that usually the extremization tasks like (\ref{theory:functional}), (\ref{theory:approx}) 
are expected to be plagued by Gribov copies problem. However, basing on the results
of Ref.~\cite{Gubarev:2005rs} we assert that this issue is most likely irrelevant.
Additional indirect evidence is provided by the fact that we changed slightly the extremization
algorithm (see below) compared to that of Ref.~\cite{Gubarev:2005rs}, however, all physically
relevant results turned out to be the same. As far as the actual numerical algorithms used in this
paper are concerned, they could be summarized as follows.
\begin{itemize}
\item $\HP{1}$ $\sigma$-model embedding.
\end{itemize}

The functional (\ref{theory:functional}) was minimized sequentially at each lattice site,
the non-linearity was taken into account by performing 3 iterations
\beq
\label{theory:iteration}
\ket{ q^{(i+1)}_x} \propto \sum\limits_\mu\left[
\frac{\ket{q_{x+\mu}} U^\dagger_{x,\mu}}{|\braket{q^{(i)}_x}{q_{x+\mu}}|} +
\frac{\ket{q_{x-\mu}} U_{x-\mu,\mu}}{|\braket{q^{(i)}_x}{q_{x-\mu}}|} \right]
\eeq
at each point $x$ and then sweeping number of times through all the lattice until
the algorithm converges.
The proportionality factor, which is omitted in (\ref{theory:iteration}), is chosen
to ensure the normalization condition $\braket{q^{(i)}_x}{q^{(i)}_x} = 1$ at each iteration.
The stopping criterion was that the distance between old and new $\ket{q}$ values at all
lattice sites in between two consecutive sweeps is smaller than $10^{-4}$
\beq
\label{theory:delta}
\max\limits_x \, ( 1 ~-~ |\braket{q^{old}_x}{q^{new}_x}|^2) ~ < ~ 10^{-4}\,.
\eeq
Note that the value of r.h.s. is the result of trade-off between
the computational demands and the needed numerical accuracy (see below).
It turns out that Eq.~(\ref{theory:delta}) provides the accuracy of $\min_{q} F(U, q)$
calculation of order $10^{-6}$  which is the same as was used in~\cite{Gubarev:2005rs}.
As far as the Gribov copies problem is concerned, we always considered 10 random initial
distributions $\ket{q_x}$ and then selected the best minimum found.

\begin{itemize}
\item Global topological charge and topological density.
\end{itemize}

The topological charge density in terms of the $\HP{1}$ $\sigma$-model fields is given
by the oriented 4-volumes of spherical tetrahedra $T$ embedded into $\HP{1} = S^4$
(image of $n^A: S^4 \to \HP{1}$). Unfortunately, the only way to estimate these
volumes is to use the Monte Carlo technique which however forbids the exact volume
evaluation. In this paper we used precisely the same algorithm of topological charge density
calculation which was discussed in length in~\cite{Gubarev:2005rs}. In particular,
the embedded $\HP{1}$ $\sigma$-model assigns 5 unit five-dimensional vectors $n^A_{{i}}$
$i=0,...,4$ to every simplex $\cT$ of physical space triangulation and the topological charge
density in simplex $\cT$ is given by
\beq
\label{theory:density}
q({\cal T}) = \frac{3}{8\pi^2} \, \mathrm{sign}\left(\mathrm{det}_{iA}\,[n^A_{(i)}]\right) \cdot V(T)\,,
\eeq
where $V(T)$ is the volume of spherical tetrahedron $T$ and
normalization factor $8\pi^2/3$ is the total volume of $S^4$. The corresponding topological charge
\beq
\label{theory:charge-float}
Q_{float} = \sum\limits_{\cal T} q({\cal T})
\eeq
is not integer valued due to finite accuracy of Monte Carlo estimation of $V(T)$.  However, for high
enough accuracy of calculation at each simplex the non-integer valuedness of the topological charge
is in fact irrelevant and we could safely identify
\beq
\label{theory:charge-int}
Q = \left[\,Q_{float}\,\right]\,,
\eeq
where $[x]$ denotes the nearest to $x$ integer number. The validity of the identification
(\ref{theory:charge-int}) was discussed in details in~\cite{Gubarev:2005rs}.
Here we note that there exist an additional cross-check of Eq.~(\ref{theory:charge-int})
based on the observation that in order to get the global topological charge
it is not necessary to calculate its density in the bulk. It is sufficient to count
(with sign) how many times a particular point $n_0 \in \HP{1}$ is covered by the image of $n^A: S^4 \to \HP{1}$
\beq
\label{theory:charge}
Q = \sum\limits_\cT \, \left\{ \begin{array}{cc}
\mathrm{sign}\,\mathrm{det}_{iA}\,[n^A_{(i)}] & \mathrm{if}~~~n_0\in T \\
0 & \mathrm{otherwise}
\end{array}\right.\,.
\eeq
Evidently, Eq.~(\ref{theory:charge}) is computationally superior to (\ref{theory:charge-int})
and we used it in present paper to calculate the topological charge. Moreover, the comparison
of (\ref{theory:charge-int}) and (\ref{theory:charge}) provides the stringent calibration of Monte Carlo
topological charge density evaluation (\ref{theory:density}).
It goes without saying that we indeed checked that both definitions
(\ref{theory:charge-int}), (\ref{theory:charge})
give exactly the same topological charge on all available configurations.
Finally, we remark that slicing of hypercubical lattice into simplices was done according
to~\cite{Kronfeld:1986ts}.

\begin{table}[t]
\centerline{\begin{tabular}{|c|c|c|c|c|c|c|} \hline
$\beta$ & $a$,fm & $L_t$ & $L_s$ & $V^{phys}, \mathrm{fm}^4$ & $N^{conf}$ & $N^{conf}_q$\\ \hline
\multicolumn{7}{l}{Fixed Volume} \\ \hline
 2.3493 & 0.1397(15) & 10 & 10 & 3.8(2)  & 300 & - \\
 2.3772 & 0.1284(15) & 14 & 10 & 3.8(2)  & 300 & - \\
 2.3877 & 0.1242(15) & 16 & 10 & 3.8(2)  & 300 & - \\
 2.4071 & 0.1164(15) & 12 & 12 & 3.8(2)  & 200 & - \\
 2.4180 & 0.1120(15) & 14 & 12 & 3.8(2)  & 100 & - \\
 2.4273 & 0.1083(15) & 16 & 12 & 3.8(2)  & 250 & - \\
 2.4500 & 0.0996(22) & 14 & 14 & 3.8(2)  & 200 & - \\
 2.5000 & 0.0854(4)  & 18 & 16 & 3.92(7) & 200 & - \\ \hline
\multicolumn{7}{l}{Fixed Spacing} \\ \hline
 2.4180 & 0.1120(15) & 14 & 14 & 6.0(3)   & 200 & - \\
 2.4180 & 0.1120(15) & 16 & 16 & 10.3(6)  & 200 & - \\
 2.4180 & 0.1120(15) & 18 & 18 & 16.5(9)  & 200 & - \\ \hline
\multicolumn{7}{l}{} \\ \hline
 2.4000  & 0.1193(9) &  16 & 16 & 13.3(4) & 198 & -  \\
 2.4750  & 0.0913(6) &  16 & 16 & 4.6(1)  & 380 & 40 \\
 2.6000  & 0.0601(3) &  28 & 28 & 8.0(2)  &  50 & -  \\ \hline
\end{tabular}}
\caption{Simulation parameters.}
\label{tab:params}
\end{table}

\begin{itemize}
\item Simulation parameters.
\end{itemize}

Our numerical measurements were performed on 14 sets (Table~\ref{tab:params})
of statistically independent SU(2) gauge configurations generated with standard Wilson action. 
The data sets are subdivided naturally into three groups indicated in Table~\ref{tab:params}.
The fixed volume configurations are exactly the same as ones used in Refs.~\cite{Gubarev:2005jm,Polikarpov:2005ey}
in studying low lying Dirac eigenmodes localization properties. This
allows us to make statistically significant comparison of $\HP{1}$ $\sigma$-model approach and
the overlap-based topological charge definition (section~\ref{section:numerics}).
Gauge configurations at fixed spacing were generated to investigate the finite volume dependence
of $\HP{1}$-based topological susceptibility  (section~\ref{section:chi}).
In addition to the gauge configurations at $\beta=2.400$, $2.475$ which were analyzed already in
Ref.~\cite{Gubarev:2005rs}, we also include the data set at  $\beta=2.600$ with
finest lattice spacing we have so far. 

The last column in Table~\ref{tab:params} represents the number of configurations on which
we calculated the bulk topological charge density using the above described algorithm.
Note that it is slightly different from that of Ref.~\cite{Gubarev:2005rs}
and this explains the relatively small number of analyzed configurations.
We stress that the results obtained with modified method remain essentially the same.
Nevertheless, for honesty reasons the old data is not included in the present study.
The relatively small number of configurations on which we calculated the bulk
topological density distribution does not spoil the statistical significance of our results.
Indeed, the majority of observables to be discussed below depend only upon the global
topological charge obtained via Eq.~(\ref{theory:charge}) and upon the corresponding
$\HP{1}$ $\sigma$-model. In turn the $\HP{1}$ fields were constructed for all configurations listed
in Table~\ref{tab:params} thus ensuring the statistical significance of our results.

The lattice spacing values quoted in Table~\ref{tab:params} were partially taken from
Refs.~\cite{Lucini:2001ej,spacings} and fixed by the physical value of SU(2) string tension
$\sqrt{\sigma} = 440~\mathrm{MeV}$.  Note that not all $\beta$-values listed in Table~\ref{tab:params}
could be found in the literature. In this case the lattice spacings and corresponding
rather conservative error estimates were obtained via interpolation
in between the points quoted in~\cite{Lucini:2001ej,spacings}.

\section{$\HP{1}$-based Topological Charge of Semiclassical and Vacuum Fields}
\label{section:numerics}

The validity of $\HP{1}$-based topological charge definition
could only be convincing provided that we confront it with other known topological
charge constructions. The preliminary study of this problem was already undertaken
in Ref.~\cite{Gubarev:2005rs}, however, only correlation of local topological densities
in $\HP{1}$ and overlap-based approaches on relatively small statistics 
was considered in that paper. Here we fill this gap and compare
the distribution of the $\HP{1}$ global topological charge $Q_{HP^{1}}$, Eq.~(\ref{theory:charge}),
with that in field-theoretical~\cite{DiVecchia:1981qi} and overlap-based~\cite{Neuberger:1997fp,overlap}
constructions.

Before presenting the results of our measurements let us discuss what could be the proper
quantitative characteristic of the correlation between various definitions of lattice
topological charge. The problem is that there is no commonly accepted viewpoint
in the literature, usually people just consider the cumulative distribution of the
topological charge in pair of definitions. For instance, in Ref.~\cite{Cundy:2002hv}
the topological charge distribution histogram in the plane~\footnote{
See below for precise definitions of both $Q_{clover}$ and $Q_{overlap}$.
} $(Q_{clover} , Q_{overlap})$ was presented, however, no quantitative measure of correlation
between $Q_{clover}$ and $Q_{overlap}$ was given. Below we discuss two observables
which quantify the notion of correlation between the topological charges $Q_i$ and $Q_j$
obtained via some abstract constructions ``i'' and ``j'' correspondingly.
The first one could be obtained from the above mentioned cumulative distribution
in $(Q_i, Q_j)$ plane by noting that the complete equivalence of ``i'' and ``j'' constructions
would give $Q_i = Q_j$ identically and therefore the entire plot would collapse into $Q_i = Q_j$ line.
On the other hand, any disagreement of ``i'' and ``j'' approaches would make the cumulative histogram
broader and therefore it is natural to consider the ratio of the widths of cumulative distribution
projected onto the lines $Q_i = - Q_j$ and $Q_i = Q_j$, for which the limiting values are
$0$ (exact equivalence of ``i'' and ``j'' constructions) and $\approx 1$ (no correlation of $Q_i$ and $Q_j$).
Moreover, it is similar to what had been done in~\cite{Cundy:2002hv} thus allowing to compare
our results with the existing literature. However, it is clear that the widths ratio
is mostly graphical characteristic since the distribution along $Q_i = -Q_j$ line is not
guaranteed to be Gaussian-like, hence the corresponding width is ambiguous in general.
To ameliorate this we propose to consider the correlator
\beq
\label{numerics:correlation}
\eta^i_j ~=~ \frac{\langle Q_i \, Q_j\rangle}{\sqrt{ \langle Q^2_i\rangle \, \langle Q^2_j \rangle }}\,,
\eeq
where averages are to be taken on the same set of configurations. Note that the normalization in
(\ref{numerics:correlation}) is such that $\eta^i_j \approx 1$ for essentially equivalent ``i'' and ``j''
constructions, while $\eta^i_j = 0$ in the case of complete decorrelation of $Q_i$ and $Q_j$.

\begin{figure}[t]
\centerline{\psfig{file=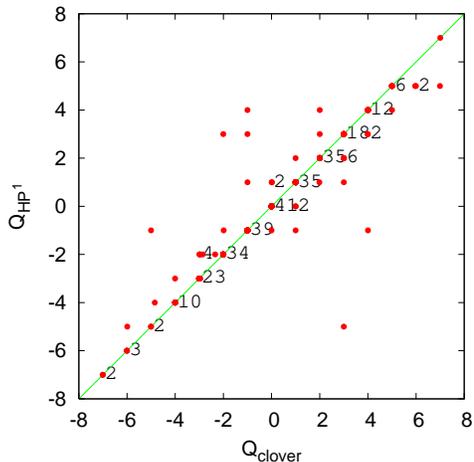,width=0.5\textwidth,silent=,clip=,angle=-90}}
\caption{Cumulative distribution of $Q_\HP{1}$, Eq.~(\ref{theory:charge})
and $Q_{clover}$, Eq.~(\ref{numerics:clover}), on 300 cooled configurations initially
thermalized at $\beta=2.400$ on $16^4$ lattice.}
\label{fig:clover-HP}
\end{figure}

With all these preliminaries, let us compare our approach with field-theoretical method,
which in turn could be applied reliably only on semiclassical (cooled) configurations.
Thus its comparison with $Q_{\HP{1}}$ is rather technical issue related mostly to the
validity of $Q_{\HP{1}}$ construction on semiclassical
background. To this end we generated 300 cooled configurations (not included
into Table~\ref{tab:params}) initially thermalized at $\beta=2.400$ on $16^4$ lattice.
The cooling algorithm employed in our study is analogous to one described in~\cite{GarciaPerez:1998ru}.
The original fields were cooled until the action stabilizes
indicating that semiclassical regime had been reached.  At this point we applied
the algorithm of $\HP{1}$ $\sigma$-model embedding thus obtaining $Q_{\HP{1}}$
and simultaneously measured the field-theoretical topological charge $Q_{clover}$ defined by
\beqn
\label{numerics:clover}
& Q_{clover} = \sum\limits_x q_{clover}(x)\,, & \\
& q_{clover}(x) = \frac{1}{32\pi^2}
\varepsilon_{\mu\nu\lambda\rho} \tr [\, F_{\mu\nu}(x) \, F_{\lambda\rho}(x)\,]\,, & \nonumber
\eeqn
where $F_{\mu\nu}$ is the $O(a^4)$ improved lattice field-strength
tensor~\cite{Bilson-Thompson:2002jk,deForcrand:1997sq}. The resulting cumulative distribution
of $Q_\HP{1}$ and $Q_{clover}$ is presented on Fig.~\ref{fig:clover-HP}.
As is evident from that figure both definitions are fairly compatible with each other
making us confident that $Q_\HP{1}$ and $Q_{clover}$ identify essentially the same topology
on all considered data sets. The disagreement between
$Q_\HP{1}$ and $Q_{clover}$ is only seen in a few points which, however, is to be expected
(see Ref.~\cite{Cundy:2002hv} for discussion of this issue
in case of field-theoretical and overlap constructions). For this reason the above discussed
widths ratio is certainly very small and is comparable with
zero within the statistical errors. As far as the correlator
(\ref{numerics:correlation}) is concerned, its numerical value is
\beq
\eta^{clover}_\HP{1} ~=~ 0.94(1)
\eeq
and indeed is very close to unity.

Turn now to the most interesting comparison of $\HP{1}$ topological charge with 
becoming standard now construction based on Atiyah-Singer index theorem applied to chirally
symmetric overlap Dirac operator~\cite{Neuberger:1997fp,overlap}.
Explicitly  the overlap operator is given by
\beq
\label{numerics:dirac}
D = \frac{\rho}{a} \left( 1 + \frac{A}{\sqrt{AA^\dagger}} \right),\,\quad
A = D_W - \frac{\rho}{a},
\end{equation}
where $A$ is the Wilson Dirac operator with negative mass term and we used  the optimal value 
$1.4$ of $\rho$ parameter. Anti-periodic (periodic) boundary
conditions in time (space) directions were employed. To compute the sign function
$\sign(A) = A/\sqrt{AA^\dagger} \equiv \gamma_5 \sign(H)$,
where $H=\gamma_5 A$ is hermitian Wilson Dirac operator,
we used the minmax polynomial approximation~\cite{Giusti:2002sm}.
In order to improve the accuracy and performance about 50 lowest eigenmodes of $H$ were projected out. 
Note that the eigenvalues of $D$ are distributed on the circle of radius
$\rho$ centered at $(\rho,0)$ in the complex plane. Below we will need to relate
them with continuous eigenvalues of the Dirac operator and therefore 
the circle was stereographically projected onto the imaginary axis~\cite{zenkin,Capitani:1999uz}.

\begin{figure}[t]
\centerline{\psfig{file=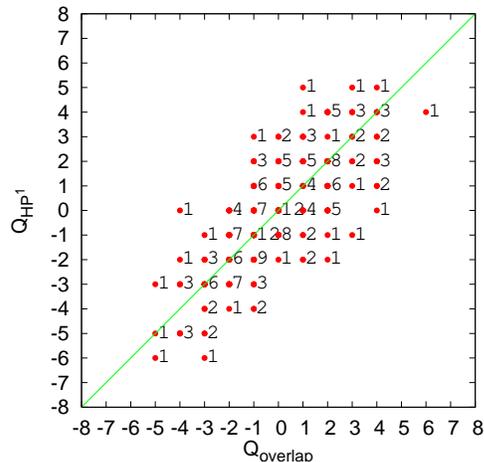,width=0.5\textwidth,silent=,clip=,angle=-90}}
\caption{Cumulative distribution of $Q_\HP{1}$, Eq.~(\ref{theory:charge})
and $Q_{overlap}$, Eq.~(\ref{numerics:overlap}), on 200 fixed volume configurations at $\beta=2.500$, Table~\ref{tab:params}.}
\label{fig:overlap-HP}
\end{figure}

\begin{figure}[t]
\centerline{\psfig{file=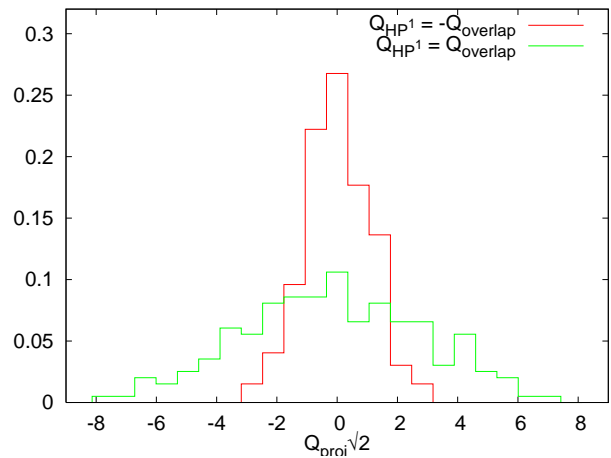,width=0.5\textwidth,silent=,clip=,angle=-90}}
\caption{Projection of the cumulative distribution of $Q_\HP{1}$ and $Q_{overlap}$ at $\beta=2.500$
onto the lines $Q_\HP{1} = \pm Q_{overlap}$, along which $Q_{proj}\, \sqrt{2}$ is the corresponding
integer valued coordinate. Note that both histograms are normalized separately.}
\label{fig:overlap-HP-histo}
\end{figure}

It is well known that the knowledge of the spectrum of (\ref{numerics:dirac}) allows to
define the topological charge via Atiyah-Singer index theorem~\cite{Atiyah:1971rm}
applied in lattice settings~\cite{overlap}
\beq
\label{numerics:overlap}
Q_{overlap} ~=~ n_+ ~-~ n_-\,,
\eeq
where $n_+$ $(n_-)$ is the number of positive (negative) chirality zero eigenmodes.
In order to compare $Q_\HP{1}$ and $Q_{overlap}$ we considered both the cumulative distribution
and the quantity (\ref{numerics:correlation})
on $200+200$ fixed volume configurations at  $\beta=2.450$ and $\beta=2.500$ (see Table~\ref{tab:params}).
Note that since the width of $Q$  distribution is generically spacing dependent it would be unwise to mix
$Q_\HP{1}$, $Q_{overlap}$ cumulative distributions at different $\beta$-values and we present
on Fig.~\ref{fig:overlap-HP} only $\beta=2.500$ results.
Evidently both definitions are strongly correlated with each other although the distribution
is notably broader than that on Fig.~\ref{fig:clover-HP}. The relative broadness of
$Q_\HP{1}$, $Q_{overlap}$ distribution is not surprising by itself, clearly it must be broader
than that on cooled configurations. What is relevant here is its projection onto the lines
$Q_\HP{1} = \pm Q_{overlap}$ and the ratio of corresponding widths. The projected histograms are presented
on Fig.~\ref{fig:overlap-HP-histo}, from which it is clear that they both are Gaussian-like,
the widths ratio being about $1/3 \div 1/4$. To get the feeling of the numbers involved
it is instructive to remind the results of Ref.~\cite{Cundy:2002hv}, where the cumulative distribution
of $Q_{clover}$ and $Q_{overlap}$ was considered. Note that the relevant plot of that paper
refers to cooled configurations and its direct comparison with Figs.~\ref{fig:overlap-HP}, \ref{fig:overlap-HP-histo}
would be unfair since for cooled configurations the distribution is generically expected to
be even narrower. However, it is pleasant to notice that the ratio of the corresponding widths
in $Q_{clover}$, $Q_{overlap}$ distribution, which could be extracted from Ref.~\cite{Cundy:2002hv},
is larger than that of $Q_\HP{1}$, $Q_{overlap}$ and is of order $1/3 \div 1/2$.
Thus we conclude that $Q_\HP{1}$, $Q_{overlap}$ correlation seems to be even stronger than
it is in the case of $Q_{clover}$ and $Q_{overlap}$ charges. Moreover, the strong correlation of
$Q_\HP{1}$ and $Q_{overlap}$ is also confirmed by the quantity (\ref{numerics:correlation})
and what is probably more important is that it is rising with diminishing spacing
\beqn
\eta^{overlap}_\HP{1}(\beta = 2.450) ~=~ 0.71(1)\,, \nonumber \\
\eta^{overlap}_\HP{1}(\beta = 2.500) ~=~ 0.77(1)\,. \nonumber
\eeqn
Note however that unlike the previous case of widths ratio the numerical values of
$\eta^{overlap}_\HP{1}$ cannot be compared with the existing literature.

To summarize, we found clear and statistically significant evidences that 
the construction of the topological charge based on $\HP{1}$ $\sigma$-model embedding 
is fairly compatible with field-theoretical definition in case of semiclassical configurations
and is strongly correlated with overlap-based approach in the case of equilibrium vacuum fields.
In the next section we put the $\HP{1}$ method under other tests, namely, consider the scaling
properties of topological susceptibility defined via $\HP{1}$ $\sigma$-model embedding.

\section{Scaling of Topological Susceptibility}
\label{section:chi}

The basic topology related observable in pure Yang-Mills theory
is the topological susceptibility which is formally defined as
\beq
\label{chi:chi-integral}
\chi ~=~ \int d^4 x \langle q(0) q(x) \rangle\,.
\eeq
Here $q(x)$ is the topological charge density which is equal to $q_{clover}(x)$,
Eq.~(\ref{numerics:clover}), in the naive continuum limit.
The topological susceptibility, being the quantity of prime phenomenological importance,
is, in fact, not well defined within the definition (\ref{chi:chi-integral})
and is written usually as
\beq
\label{chi:chi}
\chi ~=~ \lim\limits_{\stackrel{a \to 0}{\scriptscriptstyle{V \to\infty}}} \,
\langle Q^2 \rangle / V
\eeq
in the context of lattice regularization, where $a$ and $V$ are the lattice spacing and lattice volume
respectively. The purpose of this section is to study the limit (\ref{chi:chi}) and obtain the estimate
of the topological susceptibility in the continuum using the $\HP{1}$ definition of the
topological charge $Q \equiv Q_\HP{1}$.

\begin{figure}[t]
\centerline{\psfig{file=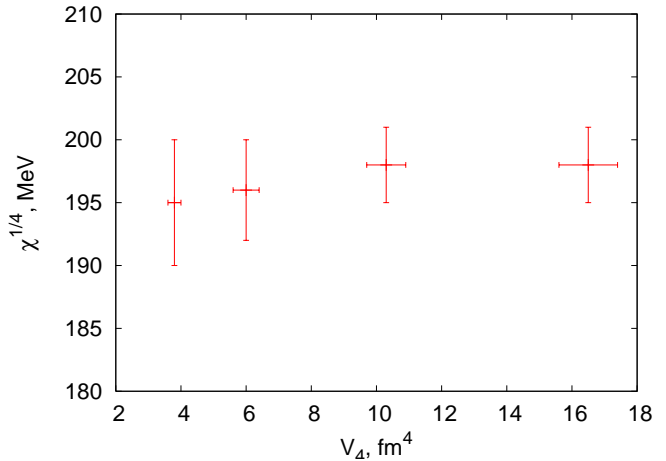,width=0.5\textwidth,silent=,clip=,angle=-90}}
\caption{The topological susceptibility as a function of lattice volume both being expressed
in physical units (fixed spacing configurations at $\beta=2.4180$, Table~\ref{tab:params}).}
\label{fig:chi-volume}
\end{figure}

\begin{figure}[t]
\centerline{\psfig{file=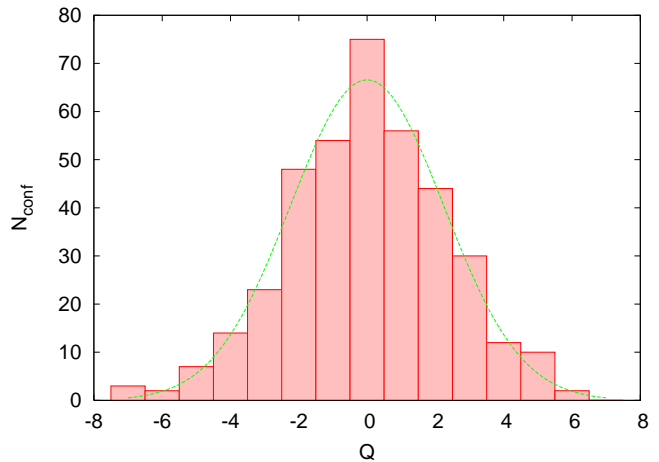,width=0.5\textwidth,silent=,clip=,angle=-90}}
\caption{Topological charge distribution obtained at $\beta=2.475$ on $16^4$ lattices
(380 configurations). Curve represents the best fit to Eq.~(\ref{chi:gauss}).}
\label{fig:Q-histo}
\end{figure}

Due to the absence of massless excitations in pure glue theory the topological susceptibility
is expected to reach its infinite volume limit exponentially fast with increasing
lattice size. The characteristic length of exponential fall off is dictated by
lightest glueball mass $(1.5 \mathrm{~GeV})^{-1} \approx 0.15 \mathrm{~fm}$
and therefore we expect that finite-volume
effects are negligible for our lattices. To confirm this assertion we measured the topological
susceptibility on fixed spacing configurations (Table~\ref{tab:params}) at $\beta=2.4180$
which cover rather wide range of lattice volumes. The result is presented on Fig.~\ref{fig:chi-volume}
and shows clearly that indeed finite size corrections are much smaller than the statistical errors.

An additional cross-check of the negligibility of finite-size corrections comes from the
observation~\cite{Giusti:2003gf,Giusti:2002sm} that the probability distribution
of the topological charge in large volume limit is to be Gaussian
\beq
\label{chi:gauss}
P_Q ~=~ \frac{1}{\sqrt{ 2\pi \langle Q^2\rangle} } \, e^{ - Q^2/(2\langle Q^2\rangle)}\,.
\eeq
We have checked that the distribution $P_Q$ is indeed as expected for all our data sets.
A particular illustration is provided by the data taken at $\beta=2.475$ on $16^4$
lattice where we have the largest statistics (Fig.~\ref{fig:Q-histo}).
Moreover, we were unable to find any statistically significant deviations of $P_Q$
from (\ref{chi:gauss}) on all our data sets.

\begin{figure}[t]
\centerline{\psfig{file=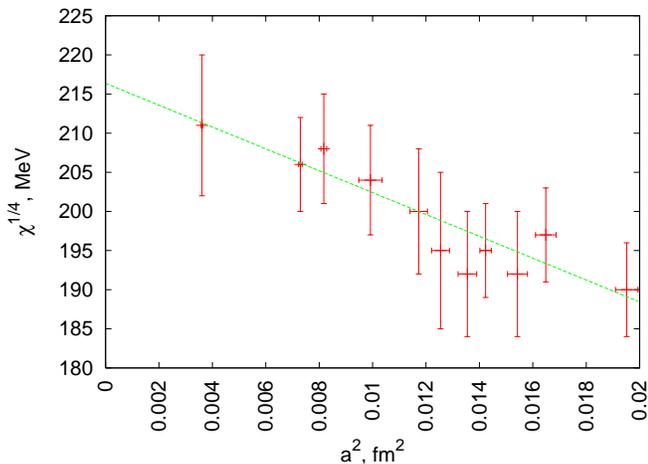,width=0.5\textwidth,silent=,clip=,angle=-90}}
\caption{Scaling of the $\HP{1}$-based topological susceptibility with lattice
spacing; line is the best fit to Eq.~(\ref{chi:fit}).}
\label{fig:chi}
\end{figure}

As far as the lattice spacing dependence of the susceptibility is concerned,
it is not {\it a priori} evident that $\chi(a)$ is not plagued by power divergences contrary
to the case of overlap-based definition. Unfortunately, we're still
lacking the theoretical arguments which could highlight the issue and
should rely, in fact, only on the results of the numerical simulations. Note however
that at least non-universal $O(a^2)$ discretization effects are expected generically. 
Therefore let us consider the topological susceptibility as a function of squared lattice spacing,
Fig.~\ref{fig:chi}. As is clear from the figure the dependence $\chi(a^2)$ is totally compatible
with linear one
\beq
\label{chi:fit}
\chi^{1/4}(a^2) = \chi^{1/4} ~+~ \alpha \cdot a^2
\eeq
and does not show any pathology in the limit of small lattice spacing.
Of course, the dependence (\ref{chi:fit}) is not guaranteed to persist in the academical limit
$a\to 0$, but our data strongly disfavors this possibility. The conclusion
is that the topological susceptibility in the $\HP{1}$ approach is not plagued by power
divergences. Moreover, the best fit according to Eq.~(\ref{chi:fit}) leads to
\beq
\label{chi:chi0}
\chi^{1/4} ~=~ 216(4) \mathrm{~MeV}\,,
\eeq
which is totally compatible with the conventional value of topological susceptibility
in the continuum limit (see, e.g., Ref.~\cite{Lucini:2001ej} and references therein).
Note that in Ref.~\cite{Gubarev:2005jm} we got $\chi^{1/4}_{overlap} = 225(3)\mathrm{~MeV}$
on essentially the same configurations using overlap-based definition of the topological charge.
We believe that this difference is completely inessential and should be attributed to the finite width
of cumulative $Q_\HP{1}$, $Q_{overlap}$ distribution along the line $Q_\HP{1} = - Q_{overlap}$,
which  we discussed in previous section.

To summarize, our high statistics computation of the topological susceptibility, which was always
considered as a testbed for all topological charge constructions, shows that the $\HP{1}$ method
is most likely to be free of any lattice related pathology. In particular, we do see almost perfect
scaling of $\chi$ within our approach and it does not reveal any sign of necessity of
multiplicative renormalization. This provides us a stringent evidence that $\HP{1}$ topological
charge is not sensitive to lattice dislocations. As far as the physical results are concerned,
the $\HP{1}$ topological susceptibility being extrapolated to the continuum limit is perfectly
consistent with the results of other investigations, most notably with overlap-based approach.
Keeping in mind all subtleties and technicalities involved in the construction of
$Q_{overlap}$ we could dare to claim that the $\HP{1}$ $\sigma$-model embedding method
is comparable with overlap based definition and is, in fact, much more advantageous computationally.
Moreover, it is not only the tool to investigate the topological aspects of equilibrium
vacuum gauge fields.
As we show below it naturally allows to study the role of topology related
fluctuations in the non-perturbative dynamics of SU(2) Yang-Mills theory.

\section{$\HP{1}$ $\sigma$-model Induced Gauge Fields}
\label{section:HP}

The essence of $\HP{1}$ $\sigma$-model embedding method is the assignment of
quaternionic valued scalar fields $\{\ket{q_x}\}$ to the original SU(2) gauge fields
configuration. The crucial question here is the uniqueness of this assignment
and, unfortunately, in the case of equilibrium gauge fields it seems to be impossible
to investigate this issue analytically. However,  the results of Ref.~\cite{Gubarev:2005rs}
and the ones presented above suggest that non-uniqueness is most likely to be
physically irrelevant. Therefore let us assume that the association
of $\{\ket{q_x}\}$ with a particular gauge configuration is indeed unique.
It is crucial that the corresponding $\HP{1}$ projection
\beq
\label{HP:projection}
U_{x,\mu} ~\to ~ U^{\HP{1}}_{x,\mu} \equiv \frac{\braket{q_x}{q_{x+\mu}}}{|\braket{q_x}{q_{x+\mu}}|}
\eeq
is gauge covariant since under gauge rotations both $U_{x,\mu}$ and $U^\HP{1}_{x,\mu}$ transform
exactly in the same way. In this respect the $\HP{1}$ projection is radically distinct from what
is usually called 'projection' in the literature. 
The very appearance of Eq.~(\ref{HP:projection}) was motivated only by consideration of the
gauge fields topology and the word 'projection' here is just the unfortunate terminology.
However, this term is in common use and we adopted it also.

Eq.~(\ref{HP:projection}) naturally allows to consider the properties of $\{U^{\HP{1}}\}$ fields
viewed as usual SU(2) matrices assigned to lattice links. The purpose of this
section is to investigate the dynamics of $\HP{1}$ induced gauge fields from various
viewpoints. In section~\ref{section:HP-plaq} we consider the simplest observable,
the gauge curvature of induced potentials, and investigate its dependence
upon the lattice spacing. Then the crucial question of confinement in the projected fields
is studied in section~\ref{section:HP-confinement}.
Finally in section~\ref{section:HP-fermions}
the properties of $\{U^{\HP{1}}\}$ configurations are investigated with chirally symmetric overlap
Dirac operator. Since this section is rather lengthy and presents quite remarkable
results, we conclude it with brief summary (section~\ref{section:HP-summary}).

\subsection{Gluon Condensate and Quadratic Corrections}
\label{section:HP-plaq}
The immediate question about $\HP{1}$ induced gauge fields is the corresponding curvature.
On the lattice this amount to the consideration of the trace of the plaquette matrix
$1/2\tr U^\HP{1}_p$ constructed as usual from $U^\HP{1}_{x,\mu}$ and it turns out
that $\HP{1}$ induced curvature is astonishingly small,
$\langle1/2\tr U^\HP{1}_p\rangle ~\gtrsim ~ 0.97$,
on all configurations we have in our disposal. To illustrate this
we note that the typical distribution of $1-1/2\tr U^\HP{1}_p$
measured on $\HP{1}$  projected configurations
is radically different from that on the original fields and is well described by power law,
which should be compared with usual almost exponential tail in $1-1/2\tr U_p$
distribution.  It is amusing that $1-1/2\tr U^\HP{1}_p$ distribution looks similar
to the lumps volume distribution reported in~\cite{Gubarev:2005rs} and most probably they are
indeed closely related. In fact, it is a clear evidence that the dynamics
of $\HP{1}$ projected fields is totally different from that of original ones although it is
still completely determined by the original Wilson action.

\begin{figure}[t]
\centerline{\psfig{file=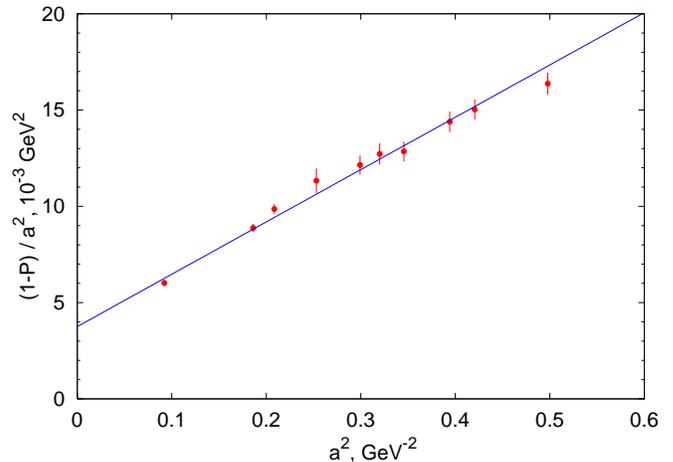,width=0.5\textwidth,silent=,clip=,angle=-90}}
\caption{Scaling of averaged plaquette $ P = \langle1/2\tr U^\HP{1}_p\rangle$
constructed from  $\HP{1}$ projected gauge fields. Line is the best fit according to Eq.~(\ref{plaq:fit}).}
\label{fig:plaq}
\end{figure}

In principle it is an interesting question how the above power law dependence changes
with lattice spacing, but we'll not directly consider it. Instead let us focus on
the spacing dependence of averaged plaquette $\langle1/2\tr U^\HP{1}_p\rangle$
for which rather accurate results could easily be obtained. We have measured the
$\HP{1}$ induced mean plaquette $\langle1/2\tr U^\HP{1}_p\rangle$ at various spacings quoted in
Table~\ref{tab:params}. The result is presented on Fig.~\ref{fig:plaq} and indeed is quite
remarkable. The spacing dependence of $\langle1/2\tr U^\HP{1}_p\rangle$ is totally distinct from
what could be expected for unprojected fields and as a matter of fact is astonishingly well
described by simple polynomial
\beq
\label{plaq:fit}
1 ~-~ \langle1/2\tr U^\HP{1}_p\rangle ~=~ \alpha_2 \, a^2 ~+~ \alpha_4 \, a^4
\eeq
throughout the whole range of spacings considered. The justify the assertion (\ref{plaq:fit})
we plot on Fig.~\ref{fig:plaq}  the dependence of $a^{-2} \cdot (1-\langle1/2\tr U^\HP{1}_p\rangle)$
upon the squared lattice spacing and it is apparent that this dependence is indeed linear.
In turn the best fit gives $\chi^2/n.d.f. = 1.2$ and the optimal values of parameters are
\beqn
\label{plaq:coef}
\alpha_2 & = & [\, 61(3) \mathrm{~MeV}\, ]^2\,, \\
\alpha_4 & = & 0.0271(10) \mathrm{~GeV}^4\,. \nonumber
\eeqn

It is quite remarkable that the dependence (\ref{plaq:fit}) shown on Fig.~\ref{fig:plaq} by solid line
includes only positive powers of lattice spacing. This means, in particular, that
$\langle1/2\tr U^\HP{1}_p\rangle$ contains no trace whatsoever of the perturbation theory
contribution and is vanishing in the limit $a \to 0$. Thus $1-\langle1/2\tr U^\HP{1}_p\rangle$
could be considered as the definition of non-perturbative contribution present in
$\langle \alpha_s\, G^2/\pi\rangle_{full}$,  where we have used the conventional continuum notations
and subscript indicates that averaging is done in the full theory including perturbative series.
We are in haste to add, however, that at the time being this definition looks completely {\it ad hoc}.
Indeed, we cannot identify $\langle 1/2\tr U^\HP{1}_p\rangle$ as the only non-perturbative
part of the $\langle \alpha_s\, G^2/\pi\rangle_{full}$ until we analyze the content
of original gauge fields not surviving $\HP{1}$ projection (\ref{HP:projection}).
We will do this in section~\ref{section:coset} and right now let us agree to treat
$\langle 1/2\tr U^\HP{1}_p\rangle$ as the genuine non-perturbative quantity.
Then it is straightforward to relate the coefficient $\alpha_4$, Eq.~(\ref{plaq:fit}),
with gluon condensate $\langle\alpha_s\,G^2/\pi\rangle$ introduced first in~\cite{Shifman:1978bx}.
Indeed, discarding temporally the $O(a^2)$ term in (\ref{plaq:fit}) we have
\beqn
1-\langle1/2\tr U^\HP{1}_p\rangle & = & \alpha_4 \, a^4(\beta) = \\
 & = & a^4(\beta) \frac{\pi^2}{12 \cdot 2} \langle\alpha_s\,G^2/\pi\rangle\,, \nonumber
\eeqn
thus obtaining the following estimation of the gluon condensate
\beq
\label{plaq:condensate}
\langle\alpha_s\,G^2/\pi\rangle ~=~ 0.066(2) \mathrm{~GeV}^4\,.
\eeq
As far as the comparison with existing literature is concerned, let us mention that the gluon condensate
being the quantity of prime phenomenological importance is most frequently discussed within
the $SU(3)$ pure gauge theory. Here the most recent lattice measurements give
$\langle\alpha_s\,G^2/\pi\rangle_{SU(3)} \approx 0.04 \mathrm{~GeV^4}$
albeit with large uncertainties (see, e.g., Ref.~\cite{Rakow:2005yn} and references therein).
This value should be compared with phenomenological one
$\langle\alpha_s\,G^2/\pi\rangle_{SU(3)} \approx 0.012 \mathrm{~GeV^4}$
coming from SVZ sum-rules~\cite{Shifman:1978bx}. We see that even in the most phenomenologically
relevant case of $SU(3)$ gauge group the estimations of the gluon condensate vary widely
and are, in fact, fairly compatible with (\ref{plaq:condensate}). If we turn now to the case
of $SU(2)$ gauge group, the results which we were able to find in the literature
are again in accordance with our estimation (\ref{plaq:condensate}) 
and vary from
$\langle\alpha_s\,G^2/\pi\rangle \approx 0.02 \mathrm{~GeV^4}$, \cite{Baig:1985pu},
to
$\langle\alpha_s\,G^2/\pi\rangle \approx 0.15 \mathrm{~GeV^4}$,  \cite{DiGiacomo:1981wt}
(for review see, e.g., Refs.~\cite{gluon} and references therein). Therefore, we are confident
that the value of the gluon condensate coming from $\HP{1}$ $\sigma$-model embedding method
agrees rather nicely with what is known in both SU(2) and SU(3) cases. Moreover,
in view of the well known tremendous uncertainties of usual approaches we conclude that
Eq.~(\ref{plaq:condensate}) provides the most accurate estimation of the gluon condensate
in $SU(2)$ gauge theory available so far. Note that the systematic errors involved in (\ref{plaq:condensate})
are coming only from $\HP{1}$ projection (\ref{HP:projection}). However, as we argue
in section~\ref{section:coset}, the systematic uncertainties are most likely to be vanishing 
thus justifying the estimate (\ref{plaq:condensate}).

Turn now to the quadratic correction term present in (\ref{plaq:fit}).
Generically it corresponds to the dimension 2 condensate
$\langle A^2_{min}\rangle$ which was introduced, in fact, long ago~\cite{A2-old}
and became the subject of active development recently~\cite{Burgio:1997hc,A2}
(see, e.g., Refs.~\cite{A2-review} and references therein). Note however that 
the theoretical status of quadratic corrections in general and,
in particular, the status $O(a^2)$ correction to the gluon condensate is uncertain at present.
Indeed, the quadratic correction to $\langle\alpha_s\,G^2/\pi\rangle$ was clearly seen in Ref.~\cite{Burgio:1997hc},
however, the corresponding coefficient decreases with increasing number of subtracted
perturbative loops~\cite{A2-plaq} and it is by no means evident whether $O(a^2)$
term has the non-perturbative origin or comes from the higher orders of perturbation
theory (see Refs.~\cite{viz,viz2} for clear and concise discussion). Generically our data (\ref{plaq:fit})
seem to disfavor the perturbative origin of the quadratic term (see also section~\ref{section:coset}).
However, as was argued in Introduction we are not in the position to interpret theoretically
this problem and would like to postpone the corresponding analyzes~\cite{future}.

As far as the actual numbers are concerned, it is remarkable that the value of $\alpha_2$ coefficient,
Eq.~(\ref{plaq:coef}), is unexpectedly small compared to the natural scale $\Lambda^2_{QCD}$,
but nevertheless almost fits into the established bounds quoted in Ref.~\cite{Rakow:2005yn}.
Note that although the data for $SU(3)$ gauge group was presented in~\cite{Rakow:2005yn},
it seems for us that the parametric smallness of the quadratic correction term is quite generic
and should also be valid in $SU(2)$ case. Therefore, we are confident that the estimate (\ref{plaq:coef})
is not in contradiction with the most recent literature.

To summarize, the measurements of the curvature of $\HP{1}$ projected
gauge fields performed in a wide range of lattice spacings reveal that these
potentials are extremely weak compared to that of the original Yang-Mills theory.
Moreover, the lattice spacing dependence of averaged $\HP{1}$ plaquette 
is astonishingly well described by simple power law, Eq.~(\ref{plaq:fit}), with only
$O(a^4)$ and $O(a^2)$ terms present and with no sign whatsoever of the perturbation theory
contribution. We argued that these terms are to interpreted
as the non-perturbative gluon condensate and unusual quadratic power correction to
$\langle\alpha_s\,G^2/\pi\rangle$. The factual absence of perturbative tail
in $\langle 1/2\tr U^\HP{1}_p \rangle$ allows us to obtain numerically precise
estimation of the gluon condensate, which agrees nicely with the existing literature.
On the other hand, the measured magnitude of the quadratic term allows to investigate
the issue of unusual power corrections in Yang-Mills theory on qualitatively new level.
As far as the systematic bias introduced by $\HP{1}$ projection (\ref{HP:projection})
is concerned, we postpone its discussion until section~\ref{section:coset}
and only note here that it is likely to be inessential.

\subsection{Confinement in Induced Gauge Fields}
\label{section:HP-confinement}
In this section we consider the crucial question of confinement in $\HP{1}$ projected
gauge fields.  In section~\ref{section:HP-confinement-wilson}
we discuss the Wilson loops confinement criterium and the scaling properties
of $\HP{1}$ projected string tension $\sigma^\HP{1}$.
Then we try to identify the objects which seem to be directly related to the non-vanishing
$\sigma^\HP{1}$. Finally, in section~\ref{section:HP-confinement-polyakov}
the Polyakov lines correlation function is considered.

\subsubsection{String Tension from Wilson loops}
\label{section:HP-confinement-wilson}
In order to investigate the confinement properties of $\HP{1}$ projected gauge fields
(\ref{HP:projection}) we have measured the planar $T \times R$ Wilson loops using various
data sets listed in Table~\ref{tab:params}. As might be expected the weakness of projected
fields allows to obtain numerically rather accurate results even without the conventional spatial
smearing~\cite{Albanese:1987ds} and hypercubic blocking~\cite{Hasenfratz:2001hp} of temporal links.
Indeed, we have checked that the utilization of both these techniques does not change our results
in any essential way.  However, both these tricks are in fact mandatory to extract the string
tension of the original gauge fields to be compared to that on $\{U^\HP{1}\}$ configurations.
In order to make the algorithms, applied to the original and $\HP{1}$ projected fields,
coherent the smearing and hypercubic blocking was done in both cases. From $\langle W(T,R) \rangle$
the heavy quark potential $V^\HP{1}(R)$ was extracted in standard way (see, e.g. Ref.~\cite{Bali:1994de} for details).
In order to obtain the $\HP{1}$ projected string tension $\sigma^\HP{1}$ the potential was fitted to
\beq
\label{pot:fit}
V^\HP{1}(R) ~=~ const ~+~ \frac{\alpha}{R} ~+~ \sigma^\HP{1} \cdot R\,.
\eeq

\begin{figure}[t]
\centerline{\psfig{file=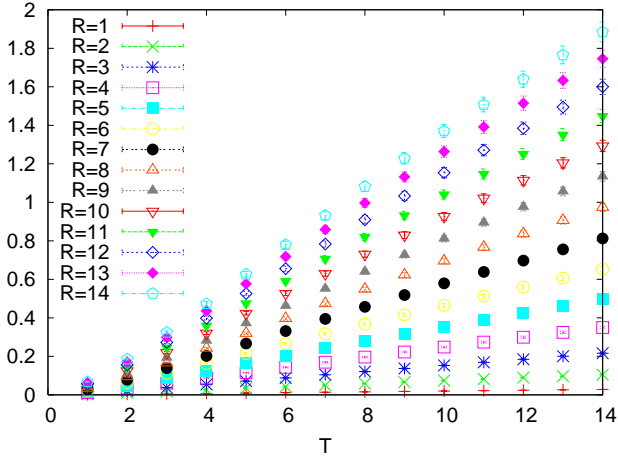,width=0.5\textwidth,silent=,clip=,angle=-90}}
\caption{$-\ln \langle W(T,R) \rangle$ as a function of $T$ at various $R$ measured at $\beta=2.600$
on $28^4$ lattice.}
\label{fig:b2.60-loops}
\end{figure}

\begin{figure}[t]
\centerline{\psfig{file=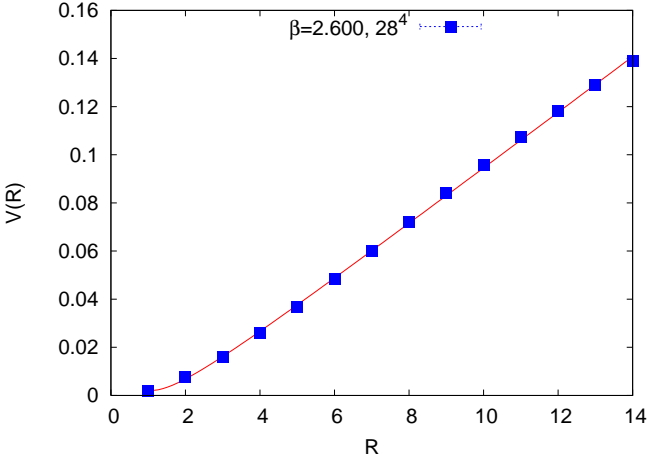,width=0.5\textwidth,silent=,clip=,angle=-90}}
\caption{Heavy quark potential $V^\HP{1}(R)$ extracted from Wilson loops measured
at $\beta=2.600$ on $28^4$ lattice. Solid curve is the fit to Eq.~(\ref{pot:fit}).}
\label{fig:b2.60-potential}
\end{figure}

To convince the reader that despite of the $U^\HP{1}$ fields weakness
the potential $V^\HP{1}(R)$ is indeed linearly rising we show on Fig.~\ref{fig:b2.60-loops}
the behavior of $-\ln \langle W(T,R) \rangle$
as a function of $T$ at various $R$ measured at $\beta=2.600$ on $28^4$ lattice.
The corresponding potential $V^\HP{1}(R)$ is depicted on Fig.~\ref{fig:b2.60-potential}. Note that the errors
bars here are smaller than the size of symbols.  It is apparent that the potential has positive second
derivative for $R < 4$ which probably indicates the reflection positivity violation. 
However, we don't think that this is a real problem. Indeed, the $\HP{1}$ projected gauge fields do not
contain the perturbative contribution (see above) and already for this reason are not obliged to fulfill
the usual requirements of reflection positivity. Additional argument comes from the intrinsic
non-locality of $\HP{1}$ projection. We could only hope to recover the refection positivity
in the large distance limit and it is pleasant to note that indeed the potential does not look
pathological for $R>3$ ($R \gtrsim 0.2 \mathrm{~fm}$ in physical units).

\begin{figure}[t]
\centerline{\psfig{file=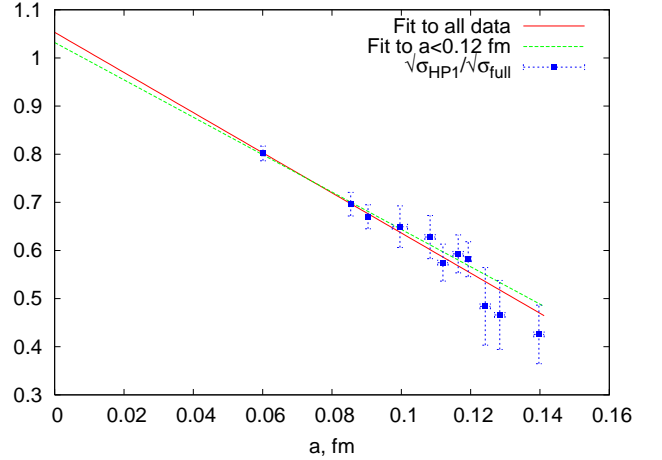,width=0.5\textwidth,silent=,clip=,angle=-90}}
\caption{Scaling of $\HP{1}$ projected string tension $\sigma^\HP{1}$ compared to that
of full $SU(2)$ string tension $\sigma^{SU(2)}$. Lines are the best fits
according to Eq.~(\ref{sigma:fit}) in the whole range of available spacings and
in the region $a < 0.12 \mathrm{~fm}$.}
\label{fig:sigma-HP1}
\end{figure}

As a matter of fact these plots are typical for all our data sets.
The only difference is that at smaller $\beta$-values we don't see any sign of the reflection
positivity violation at small $R$, the potential is everywhere linear within the errors bars. However, the projected
string tension  depends non-trivially on the lattice spacing. 
In order to investigate $\sigma^\HP{1}$ scaling properties we calculated the full $SU(2)$
string tension $\sigma^{SU(2)}$ on our configurations and considered the ratio of
$\sqrt{\sigma^\HP{1}}$ and $\sqrt{\sigma^{SU(2)}}$. This trick allows us not only to cross-check
the physical scale of our fixed volume configurations, but also to avoid the finite volume
systematic uncertainties. The resulting plot of the ratio $[\sigma^\HP{1}/\sigma^{SU(2)}]^{1/2}$
with rather pessimistic error estimates as a function of lattice spacing is presented on
Fig.~\ref{fig:sigma-HP1}. It is clear that the scaling of string tensions ratio is fairly compatible with
linear $a$-dependence
\beq
\label{sigma:fit}
\sqrt{\sigma^\HP{1}} / \sqrt{\sigma^{SU(2)}} ~=~  c_0 ~+~ c_1 \cdot a\,.
\eeq
Indeed, on general grounds we could exclude the possibility
of negative powers of $a$ in (\ref{sigma:fit}) while our data is practically insensitive
to the inclusion of $O(a^2)$ terms. Apparently three points corresponding to the largest spacings
systematically deviate from (\ref{sigma:fit}) which can be attributed the closeness of the crossover
transition. Therefore we could try to fit the data to Eq.~(\ref{sigma:fit}) either in the region
$a < 0.12 \mathrm{~fm}$ or in the whole range of available spacings. It turns out that $c_0$ value
is practically independent upon the concrete choice of the fitting range being
$1.03(2)$ and $1.05(2)$ in both cases correspondingly. Therefore the value of the string tensions ratio  in the continuum
limit is
\beq
\label{sigma:continuum}
\sqrt{\frac{\sigma^\HP{1}}{\sigma^{SU(2)}}} ~=~  1.04(3)\,.
\eeq
Note that this result is indeed remarkable. Taken at face value it indicates that in the continuum
limit the full $SU(2)$ string tension arises solely due to the rather weak $\HP{1}$ projected
fields. Simultaneously this is the first indication that what had been cut away from 
the original fields by $\HP{1}$ projection corresponds likely to pure perturbation theory.
On the other hand, Eq.~(\ref{sigma:continuum}) should not come completely unexpected.
Indeed, the non-vanishing value of the dimension 2 condensate in $\HP{1}$ projected fields
(see section~\ref{section:HP-plaq}) is a hint that the projected theory might be
confining~\cite{viz,viz2,A2-confine}.
However, we will not dwell on this issue any longer but consider the anatomy of $\HP{1}$
string tension instead.

\subsubsection{String Tension and Topological Fluctuations}
\label{section:HP-confinement-lumps}
One of the central point of Ref.~\cite{Gubarev:2005rs}
was rediscovery of the lumpy structure~\cite{Gattringer,DeGrand,Hip,Horvath}
of the topological charge density bulk distribution in $\HP{1}$ $\sigma$-model approach.
As was stressed in that paper, the term 'lumps' is in fact uncertain
until one introduces a particular cutoff $\Lambda_q$ on the magnitude of the local topological charge density.
Indeed, the most straightforward argument here is that in the numerical simulations the
topological density is always known with finite accuracy and therefore a particular cutoff
applies inevitably. Moreover, the introduction of finite $\Lambda_q$ is inherent
to practically all studies of the gauge fields topology. For instance, the overlap-based topological
density, which is given by the sum of Dirac eigenmodes $\psi_\lambda$ contributions,
is usually either restricted to lowest modes, $\lambda < \Lambda$, or is considered for all modes
available on the lattice, $\lambda \lesssim 1/a$. In both cases it is apparent that a particular 
cut on the topological density is introduced although it might not be simply expressible
in terms of $\Lambda$ or $1/a$.  Another example of this kind is provided by the investigations of
local chirality of Dirac eigenmodes~\cite{Cundy:2002hv,Gattringer,DeGrand,Hip,Horvath}.
Here the local chirality is considered only in points at which the eigenmode is reasonably large.

Then the crucial question is the dependence of physical observables
upon $\Lambda_q$. The point of view accepted in Ref.~\cite{Gubarev:2005rs} is that
this dependence must be trivial since $\Lambda_q$ is technical rather than physical parameter.
This requirement fixes, in fact, the physically meaningful values of $\Lambda_q$ ('physical window'
in the terminology of Ref.~\cite{Gubarev:2005rs}) which appears to be rather narrow and is around
$[ 300-350 \mathrm{~MeV}]^4$. The attempt to consider the limit $\Lambda_q\to 0$
results in rather peculiar structure of topological excitations which is reminiscent to one
discovered in~\cite{Horvath-structures} with overlap-based definition of the topological density.
Namely, the density of lumps seems to be divergent with $\Lambda_q\to 0$ due to the abundance of the lumps
consisting of only one lattice site. Simultaneously lumps of larger volumes are accumulating
in a few (typically two) percolating lumps with linear extent equal to the size of the lattice.
However, both the small lumps distribution and the volume occupied by percolating lumps
are strongly $\Lambda_q$-dependent and for that reason the small values of the cutoff were
excluded in Ref.~\cite{Gubarev:2005rs} from the $\Lambda_q$ physical window.
It is important that the topological susceptibility remains lumps-saturated even
beyond the above percolation transition and essentially this was the justification
for the existence of $\Lambda_q$ physical window.

It is natural then to ask the same question about the $\HP{1}$ string tension, namely,
to consider $\Lambda_q$ dependence of $\sigma^\HP{1}$, Eq.~(\ref{pot:fit}).
Evidently we should confront the $\Lambda_q$ physical window found in~\cite{Gubarev:2005rs}
with the requirement of $\Lambda_q$ independence of $\sigma^\HP{1}$. The first problem to be addressed
here is how the string tension could at all be dependent upon the cutoff on the topological density.
For the topological susceptibility the solution is trivial, it suffice to consider the small
values of the topological density $|q(x)| < \Lambda_q$ as being zero exactly.
Now we have to formulate the same principle at the level on $\HP{1}$ projected gauge fields
maintaining the gauge covariance. The straightforward way to achieve this is provided by
Eq.~(\ref{theory:inhomo}). Indeed, every $\HP{1}$ induced link is representable as
\beqn
\label{lumps-representation}
U^\HP{1}_{x,\mu} & = & \frac{\braket{q_x}{q_{x+\mu}}}{|\braket{q_x}{q_{x+\mu}}|} = \\
 & = & \bar g_x \frac{[ 1 ; \bar\omega_x] \, [ 1 ; \omega_{x+\mu} ]^T}{| 1 + \bar\omega_x \omega_{x+\mu}|} g_{x+\mu} 
\equiv \bar g_x \, h_{x,\mu} \, g_{x+\mu}\,, \nonumber
\eeqn
where $h_{x,\mu} \in SU(2)$ is, in fact, gauge invariant quantity.
If the topological density in the vicinity of point $x$ is small, $|q(x)| < \Lambda_q$, it is sufficient
to equate the inhomogeneous coordinates $\omega_x$, $\omega_{x+\mu}$ to each other  thus replacing
\beq
\label{lumps-cutoff}
h_{x,\mu} ~\to~ 1\,, \qquad U^\HP{1}_{x,\mu} ~ \to ~ \bar g_x \, g_{x+\mu}\,.
\eeq
This receipt is indeed natural since it maintains gauge covariance and nullifies the contribution
of the link $U^\HP{1}_{x,\mu}$ to any Wilson loop. Note however that we are not allowed to change
the links and topological density at other points. This means that Eq.~(\ref{lumps-cutoff})
should be understood on the level of links only, it is not allowed to actually change neither
$\ket{q_x}$ nor $\ket{q_{x+\mu}}$. Our implementation of this approach is as follows:
the $U^\HP{1}_{x,\mu}$ link matrix is transformed according (\ref{lumps-cutoff})
if the magnitude of the topological density in 8 neighboring hypercubes containing
the link $\{x,\mu\}$ is smaller than $\Lambda_q$.

\begin{figure}[t]
\centerline{\psfig{file=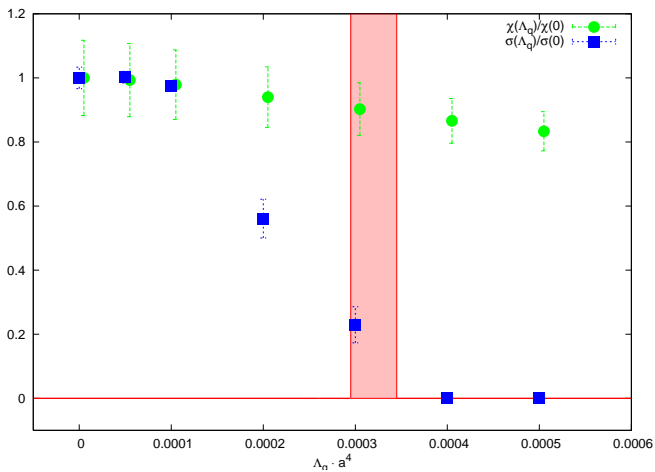,width=0.5\textwidth,silent=,clip=,angle=-90}}
\caption{$\Lambda_q$ dependence of the $\HP{1}$ string tension (squares) and of the topological susceptibility
(circles) both normalized to their respective values at $\Lambda_q = 0$ ($\beta=2.475$, $16^4$).
Shaded box stands for the approximate location of the lumps percolation transition (see text).}
\label{fig:sigma-lumps}
\end{figure}

Unfortunately, the only set of configurations on which we could analyze $\Lambda_q$ dependence
of $\sigma^\HP{1}$ is the one corresponding to $\beta=2.475$ on $16^4$ lattice, thus no scaling checks
are possible at the time being. The $\Lambda_q$ dependence of the ratios
$\sqrt{\sigma^\HP{1}(\Lambda_q)/\sigma^\HP{1}(0)}$ and $\sqrt{\chi(\Lambda_q)/\chi(0)}$, where
$\sigma^\HP{1}(0)$  and $\chi(0)$ are the $\HP{1}$ string tension and topological susceptibility
with no cutoff imposed, is presented on Fig.~\ref{fig:sigma-lumps}.
It is remarkable that the sensitivity of these  quantities to the $\Lambda_q$ cutoff are radically distinct.
The susceptibility decreases only slightly with rising $\Lambda_q$ and is changed by $\approx 20\%$
in the whole range of cutoff considered. Contrary to that the $\HP{1}$ string tension drops
down to zero at $\Lambda_q \cdot a^4 \approx 3 \cdot 10^{-4}$
and remains vanishing for all higher values of the cutoff. Moreover, we cannot exclude
the possibility of the phase transition in the infinite volume limit.
It is crucial that the string tension vanishes
just below the $\Lambda_q$ physical window
$\Lambda_q \cdot a^4 \in (4 \div 8 ) \cdot 10^{-4}$ discussed in
Ref.~\cite{Gubarev:2005rs}. In turn this is not surprising since the $\HP{1}$ string tension
was not considered in that paper.

In fact, it is not difficult to identify what happens physically around the point 
$\Lambda_q \cdot a^4 \approx 3 \cdot 10^{-4}$. Indeed, exactly around this value of $\Lambda_q$
the percolating lumps cease to exist, the gap separating percolating lumps from the remaining spectrum
in the lumps volume distribution disappears. To illustrate this point we indicated the approximate
location of the lumps percolation transition by shaded box on Fig.~\ref{fig:sigma-lumps}.
Unfortunately, at present we are not able formulate
this qualitative picture more rigorously, mainly because of the limited data we have right now.
Nevertheless, we do believe that the disappearance of $\HP{1}$ string tension could be formulated
equivalently in the language of lumps percolation transition. The corresponding strong correlation
is evident already on our limited data sets and it is probably the question of relatively short time to make
this conjecture numerically convincing~\cite{future}.

To summarize, the investigation of $\HP{1}$ string tension dependence on the cutoff imposed
on the topological charge density reveals that the string tension is most likely due
to the prolongated (extending through all the lattice) regions of sign-coherent topological charge
first discovered in~\cite{Horvath-structures}
and discussed within $\HP{1}$ approach in~\cite{Gubarev:2005rs}.
Therefore the string tension in $\HP{1}$ projected configurations, which accounts for the full
$SU(2)$ string tension in the continuum limit, is most sensitive not to the magnitude
of the topological density fluctuations, but rather to the long range correlations between them.
Contrary to that the topological susceptibility is almost blind to such correlations
and, in fact, is saturated by almost random lumps in the topological density bulk distribution.

\subsubsection{Polyakov lines correlation function}
\label{section:HP-confinement-polyakov}
Let us discuss another conventional confinement indicating observable, namely,
the Polyakov line and the corresponding correlation function. The basic observation here
is that the Polyakov line ceases to be an order parameter once the gauge potentials
(or link matrices in lattice settings) are written as
\beq
\label{pol:representation}
A_\mu ~=~ \bra{q}\diff_\mu \ket{q}\,, \qquad 
U_{x,\mu} ~=~ \frac{\braket{q_x}{q_{x+\mu}}}{|\braket{q_x}{q_{x+\mu}}|}\,.
\eeq
The intrinsic reason for this is that the relevant global center symmetry cannot be formulated
in terms of $\ket{q_x}$ fields. Hence the expectation value $\langle P \rangle$ of the Polyakov line
is not obliged to vanish in confinement phase once the representation (\ref{pol:representation}) is adopted.
Physically this is because it is always possible to form the gauge invariant
composite state $\ket{q_x} \psi_x$  from $\ket{q}$ field and quark operator $\psi$.
More explicitly, consider for instance the derivation presented Ref.~\cite{McLerran:1981pb}.
The equation which is crucial here is the static time-evolution equation
\beq
\label{pol:static}
(\diff_t ~+~ A_0 ) \, \psi(t, \vec{x})  ~=~  0\,,
\eeq
which eventually produces the Polyakov line gauge transporter to be averaged in the thermal ensemble
(note that our gauge potentials are anti-hermitian, hence there is no imaginary unit in (\ref{pol:static})).
Within the representation (\ref{pol:representation}) this equation is equivalent to
(we omit the inessential $\vec x$ dependence here)
\beq
\bra{q_t} \, \diff_t [\, \ket{q_t}\,\psi_t\, ] ~=~ 0\,,
\eeq
which could be solved as
\beq
\psi_t = \bra{q_t} \, \int\limits_0^t [\, \ket{q^\perp} \cdot f \,] ~+~ \braket{q_t}{q_0} \cdot \psi_0 \,,
\eeq
where the state $\ket{q^\perp}$ is orthogonal to $\ket{q}$, $\braket{q}{q^\perp} = 0$, and is determined,
in fact, uniquely, $f$ is arbitrary function and the second term on the r.h.s. is determined from
boundary condition at $t=0$. Evidently, for single quark free energy $F_q$,
considered in the canonical formalism, only the last term does matter
(see Ref.~\cite{McLerran:1981pb} for details) and we obtain
\beqn
\label{pol:screen}
e^{-F_q/T} & = & \tr[\, e^{-H/T} \, \psi_{1/T}\, \psi^\dagger_0\,] = \\
 & = & \tr[\, e^{-H/T} \, \braket{q_{1/T}}{q_0}\,] = 1\,, \nonumber
\eeqn
where $H$ is the Hamiltonian, $T$ is the temperature and
we assumed that the periodic boundary condition in imaginary time are imposed on the $\ket{q}$ fields.
The result is in agreement with what was noted above: the color charge of both
$\psi_{1/T}$ and $\psi_0$ gets screened by the quanta of $\ket{q}$ fields.
Note that Eq.~(\ref{pol:screen}) is not dynamical, it is valid irrespectively at both low and
high temperatures. Moreover, it confirms that the Polyakov line expectation value $\langle P \rangle$
measured on $\HP{1}$ projected fields is not related to the free energy of static color source
and generically is not vanishing at low temperatures
\beq
\label{pol:expectation}
\langle P \rangle ~\ne ~ 0\,.
\eeq
In fact, Eq.~(\ref{pol:expectation}) is strictly confirmed by our measurements. Depending on the lattice
size and spacing $\langle P \rangle$ varies from $\approx 0.42$ to $\approx 0.67$
revealing non-trivial dependence on both $V$ and $a$. However, we did not investigated this dependence
due to the uncertainty of its physical implications. On the other hand, the physical meaning
of the Polyakov lines connected correlation function is not changing
\beq
\label{pol:correlator}
\langle P_0 \, P_x \rangle - \langle P \rangle^2 ~\sim~
\exp\{\, - \frac{1}{T}\,V^\HP{1}(|x|)\,\}\,.
\eeq
Due to the weakness of $\HP{1}$ projected fields the behavior (\ref{pol:correlator})
is firmly confirmed even on the largest lattices we have. It turns out that the potential
$V^\HP{1}(R)$ extracted from (\ref{pol:correlator}) is in agreement with what we have discussed
in section~\ref{section:HP-confinement-wilson} 
although it has much larger error bars due to the subtraction of disconnected contribution.
In particular, it approaches linear asymptotic at large distances, the corresponding string
tension being in agreement with that of section~\ref{section:HP-confinement-wilson}.

\subsection{Chiral Fermions in $\HP{1}$ Projected Gauge Background}
\label{section:HP-fermions}
It had long been understood that the problem of confinement and chiral symmetry breaking
are closely related~\cite{Gross:1974jv,fermions-confinement}
(for recent discussions see, e.g., Refs.~\cite{fermions-reviews}).
In this section we investigate the properties of $\HP{1}$ projected fields (\ref{HP:projection})
as they are seen by fermionic probes provided by the overlap Dirac operator (\ref{numerics:dirac}).
Physically the most important part of the spectrum of (\ref{numerics:dirac}) is given by low lying eigenmodes
\beq
D \, \psi_\lambda ~=~ \lambda \, \psi_\lambda\,,
\eeq
the spectral density $\rho(\lambda)$ of which gives the quenched chiral condensate
via Banks-Casher relation~\cite{banks-casher}
$\langle\bar\psi\psi\rangle ~=~ \pi \lim_{\lambda \to 0} \rho(\lambda)$.
Another problem of prime importance is the localization properties of low lying modes,
which is conventionally formulated in terms of the inverse participation ratio (IPR) $I_\lambda$
(see, e.g., Refs.~\cite{Gubarev:2005jm,Polikarpov:2005ey,IPR-defs} for definitions).
Specifically, we are interested in the scaling of fermionic IPRs with lattice spacing
and will not consider dependence upon the total volume (see Refs.~\cite{Greensite:2005dv}
for the similar treatment of scalar probe particles).
We study both $\rho(\lambda)$ and $I_\lambda$ on $\HP{1}$ projected gauge fields obtained
from our fixed volume data sets (Table~\ref{tab:params}) which cover rather wide range of
lattice spacings and allow to investigate the scaling of $\rho(\lambda)$ and $I_\lambda$ in details.
Note that the spacing dependence of IPR for near zero eigenmodes is
still the subject of intensive discussions in the literature~\cite{Gubarev:2005jm,Polikarpov:2005ey,IPR-defs}.
Thus the $\HP{1}$ $\sigma$-model embedding method again provides the unique opportunity to study the issue.

One technical note is now in order. To speed up the calculations we considered only 20
lowest Dirac eigenmodes and excluded the exact zero modes from the analysis since
zero modes are known to be physically quite different from the remaining part of the spectrum.
This means that we were able to collect the sufficient statistics only for 
$\lambda \lesssim 400\mathrm{~MeV}$ and for this reason all the graphs below are
snipped off for higher eigenvalues. However, this does not spoil the
significance of our results since exactly the region $\lambda \lesssim 400\mathrm{~MeV}$
is most important physically.

\begin{figure}[t]
\centerline{\psfig{file=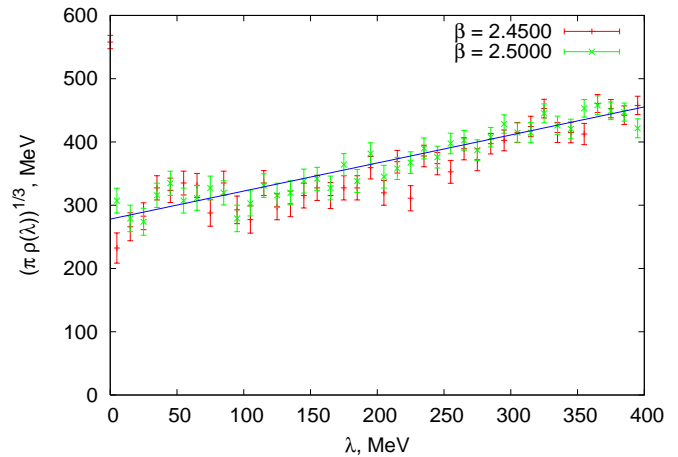,width=0.5\textwidth,silent=,clip=}}
\caption{The spectral density of lowest Dirac eigenmodes calculated on fixed volume configurations,
Table~\ref{tab:params}, at different spacings. Line represents linear interpolation towards $\lambda = 0$.}
\label{fig:HP-spectrum}
\end{figure}

Let us consider first the spacing dependence of the spectral density $\rho_\lambda$ at small
$\lambda$ as is measured on $\HP{1}$ projected gauge fields, Fig.~\ref{fig:HP-spectrum}.
Note that for readability reasons not all available graphs are depicted since the data
sets being expressed in physical units look essentially the same at different spacings.
It is clear that the points at various $a$ are falling, in fact, on the top of each other
thereby confirming that the spectral density of low lying modes is spacing independent.
This not only justifies the assertion made earlier that the $\HP{1}$ projection captures correctly
the topological aspects of the original gauge fields, but also allows to estimate the quenched chiral condensate
\beq
\langle\bar\psi\psi\rangle ~=~ \pi \lim_{\lambda \to 0} \rho(\lambda) ~=~ [278(6)\mathrm{~MeV}]^3\,,
\eeq
which is in agreement with its value measured on the same original gauge configurations
in Ref.~\cite{Gubarev:2005jm}
and is fairly compatible with magnitude of $\langle\bar\psi\psi\rangle$
known in the literature (see, e.g., Refs.~\cite{psi-psi} and references therein).

\begin{figure}[t]
\centerline{\psfig{file=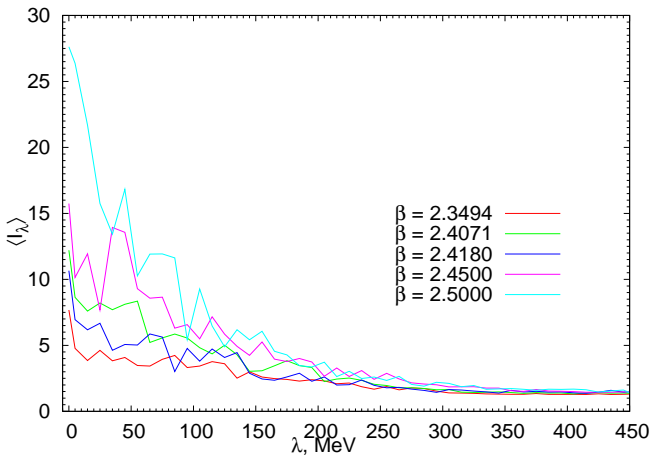,width=0.5\textwidth,silent=,clip=}}
\caption{IPRs $I_\lambda$ for lowest eigenmodes at various lattice spacings. Note the explicitness
of the 'mobility edge' around $\lambda \approx 200\mathrm{~MeV}$.}
\label{fig:HP-IPR}
\end{figure}

\begin{figure}[t]
\centerline{\psfig{file=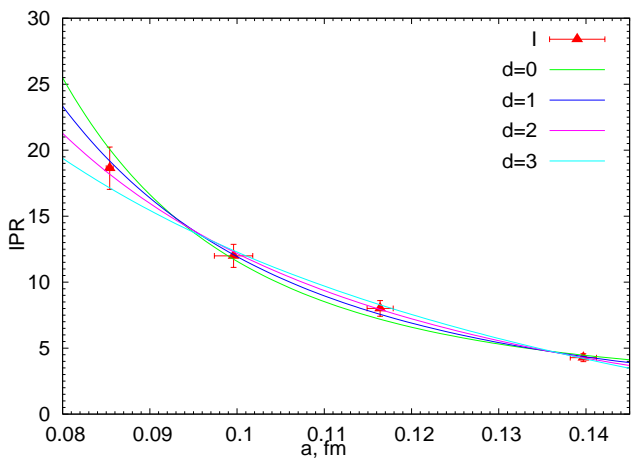,width=0.5\textwidth,silent=,clip=}}
\caption{IPR for lowest eigenmodes, Eq.~(\ref{fermi:IPR}), as a function of lattice spacing.
The fitting curves (\ref{fermi:fit}) are represented by various lines (see text).}
\label{fig:HP-localization}
\end{figure}

As far as the localization properties of lowest eigenmodes are concerned, our results
are qualitatively similar to ones reported in~\cite{Gubarev:2005jm} (see also~\cite{Polikarpov:2005ey,IPR-defs}).
Namely, for $\lambda \lesssim 200 \mathrm{~MeV}$ the eigenfunctions show notable localization
which is apparent from Fig.~\ref{fig:HP-IPR},
the corresponding IPRs are ranging from $4\div 8$ at $\lambda \approx 150\mathrm{~MeV}$ to
$10\div 25$ at smallest non-zero eigenvalues. However, the most important problem is
the spacing dependence of $I_\lambda$, to study which it is advantageous to consider
the IPR averaged over all lowest modes $0 < \lambda < \Lambda = 50 \mathrm{~MeV}$
\beq
\label{fermi:IPR}
I ~=~ \frac{1}{N_\Lambda} \sum_{\lambda < \Lambda}  \, I_\lambda\,,
\eeq
where $N_\Lambda$ is the total number of modes in the above interval. This allows to
notably improve the statistics, however,  we checked that the  scaling properties of $I$
are $\Lambda$-independent provided that $\Lambda \lesssim 100\mathrm{~MeV}$. 
The lattice spacing dependence of $I$ allows to investigate quantitatively the dimensionality
$d$ of the regions which localize the lowest eigenmodes. As it is customary now (see, e.g.,
Refs.~\cite{Gubarev:2005jm,Polikarpov:2005ey,IPR-defs} for details) we fitted the $I(a)$ dependence to
\beq
\label{fermi:fit}
I(a) ~=~ b_0 ~+~ b_1 \cdot a^{d-4}\,,
\eeq
the best fitting curves for $d=0,1,2,3$ are shown by lines on Fig.~\ref{fig:HP-localization}.
Unfortunately, our data does not allow to pinpoint the dimensionality $d$ with high enough accuracy.
However, it still favors the two-dimensional structure of the localization regions although
we cannot fairly exclude $d=1$ possibility.  The relevant $\chi^2/n.d.f.$  is equal to
$0.33$ for $d=2$ and $0.62$ for $d=1$, while being significantly larger in $d=0$ and $d=3$ cases
($1.26$ and $0.82$ correspondingly).
In either case the data is incompatible with naively expected localization on four-dimensional
'balls' of finite physical extent, the localization volumes are definitely shrinking in the limit $a\to 0$,
which is in accord with what is known in the literature~\cite{Gubarev:2005jm,Polikarpov:2005ey,IPR-defs}
about the low lying fermionic eigenmodes in equilibrium gauge background.

\subsection{Summary}
\label{section:HP-summary}
Let us briefly summarize the results obtained so far on the dynamics of
$\HP{1}$ projected gauge fields. We stress from very beginning that the $\HP{1}$ projection is completely
different from what is usually known by the term 'projection' in the literature. Namely,
it maintains the gauge invariance and in view of the investigated irrelevance
of Gribov copies problem provides unique gauge invariant variables characterizing
the original gauge background.
The main conjecture of this section to be justified below is that $\HP{1}$ projected
fields represent almost entirely the non-perturbative content of the original gauge fields.
We illustrated this point from various perspectives starting from the simplest possible
observable, namely, the $\HP{1}$ induced gauge curvature. Remarkably enough, the projected
fields appear to be extremely weak, the averaged plaquette $\langle 1/2\tr U^\HP{1}_p\rangle$ being
distributed with striking power law, which is to be compared with usual exponential distribution.
As a consequence, the lattice spacing dependence of $\langle 1/2\tr U^\HP{1}_p\rangle$ 
is totally different from what could be usually expected, we found no sign whatsoever
of the perturbation theory in it. As a matter of fact $ 1 - \langle 1/2\tr U^\HP{1}_p\rangle$
vanishes in the limit $a\to 0$, the leading spacing corrections being of order $O(a^4)$ and $O(a^2)$ only.
The corresponding coefficients are to be identified with the gluon 
condensate and quadratic power correction, both of which we measured
with highest accuracy available so far. As far as the gluon condensate is concerned, 
its value is in a complete agreement with the existing literature.
At the same time, the numerically precise value of the quadratic correction given in lattice spacing
normalization is reported for the first time and allows to pose the problem of
unusual power corrections in Yang-Mills theory on qualitatively new level.

Then we addressed the problem of confinement in the $\HP{1}$ projected gauge fields.
Despite of the relative weakness of the $\HP{1}$ induced potentials we showed that
the projected theory is still confining. Although at finite lattice spacing
the projected string tension is smaller than that on the original fields, it nevertheless
reproduces the full $SU(2)$ string tension in the limit $a\to 0$. Moreover, the last
statement was found to be quite robust with respect to the technicalities involved. 
However, it would be plainly wrong to rely on the $\HP{1}$ projection at all distances in particular
because the method
is practically blind to the perturbation theory. It turns out that the heavy quark potential at small
$R \lesssim 0.2 \mathrm{~fm}$ scales clearly violates the reflection positivity requirements,
which, however, is to be expected due to the intrinsic non-locality of $\HP{1}$ $\sigma$-model embedding.
We also argued that within our approach the usual confinement sensitive observables like
Polyakov line expectation value $\langle P \rangle$ is to be considered with care. In particular,
the notion of global center symmetry, which is crucial to ensure $\langle P \rangle = 0$
at low temperatures in usual settings, does not exist within the $\HP{1}$ projection method.
This clearly rises the question on the relevance of the center symmetry to confinement,
however, we postponed the corresponding discussions~\cite{future} and concentrated only
on the experimental results instead.

Since the $\HP{1}$ projection is ultimately related to the topology of the gauge fields,
it makes possible the investigation of the microscopic origin of the projected string tension.
We found that the global (percolating) regions of sign-coherent topological
charge~\cite{Horvath-structures,Gubarev:2005rs} seem to be fundamentally important.
Namely, the geometrical explicitness of $\HP{1}$ $\sigma$-model, being superior to any other
known topology considerations, allows to scan the gauge fields topology at various scales and 
to associate rather unambiguously the projected string tension with the existence of the above regions.
Note that at present much more data is needed to put the issue on the numerically solid grounds, but
for us the correlation is too much evident to be in any doubts.

Finally, in order to present the comprehensive picture of the projected fields dynamics
we considered the spectrum of the overlap Dirac operator in the $\HP{1}$ induced background.
The observables here are the spectral density of low lying eigenmodes and their
localization properties conventionally encoded into inverse participation ratio (IPR).
We found that both this quantities are essentially the same as they were on unprojected fields
although the actual measurements are much simpler due to the weakness of $\HP{1}$ potentials.

However, the above conclusions rely strongly on the fact that $\HP{1}$ projection leaves
aside only the perturbation theory. It is the purpose of the next section to demonstrate
that indeed what had been left after the $\HP{1}$ projection likely corresponds to purely
perturbative fields.

\section{What had been Cut Away by $\HP{1}$ Projection?}
\label{section:coset}
The problem to be considered below is the properties of the fields $V_{x,\mu}$
which are the part of original configurations cut away by $\HP{1}$
projection (\ref{HP:projection}). Symbolically $V_{x,\mu}$ could be defined
by $V = U / U^\HP{1}$, however, this equation makes no sense as it stands. 
Indeed, both $U$ and $U^\HP{1}$ transform as usual under the gauge transformations
and hence the transformation properties of $V$ are undefined.
It is clear that we could make sense of $V$ only in a particular gauge,
however, we are not aware of any detailed treatment of this generic problem in the lattice context.
The only exception is probably Ref.~\cite{Bornyakov:2005wp} where related issues are discusses.
Although the extraction of $V$ variables seems to be similar to the background fields technique,
we note that usually the background potentials are taken for granted already in a particular gauge
thus invalidating any similarities with the present case.

\begin{figure}[t]
\centerline{\psfig{file=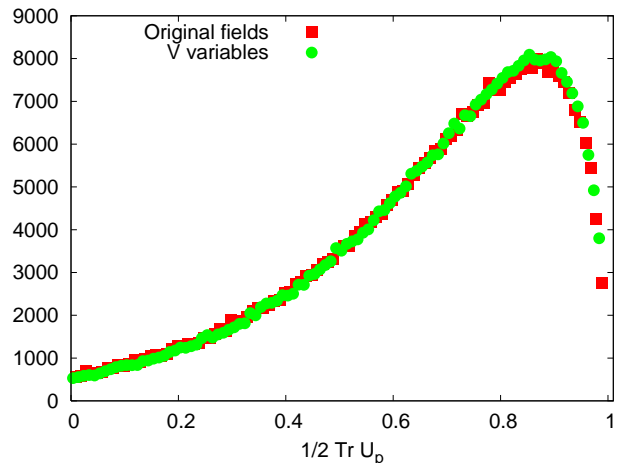,width=0.5\textwidth,silent=,clip=,angle=-90}}
\caption{Distribution of trace of the plaquette matrix constructed for a particular configuration
($\beta=2.475$, $16^4$) from $U_{x,\mu}$ and $V_{x,\mu}$ fields.}
\label{fig:coset-plaq}
\end{figure}

In order to provide the unambiguous meaning to $V_{x,\mu}$ fields we fixed the most convenient and well defined
gauge, in which all $\ket{q_x}$ variables are given by the r.h.s. of Eq.~(\ref{theory:inhomo})
\beq
\label{coset:gauge-fixing}
\ket{q_x} ~ \to ~ \frac{1}{\sqrt{1 + |\omega_x|^2}} \cdot [\,1\, ;\, \omega_x\,]^T\,.
\eeq
Then $V_{x,\mu}$ variables are defined by
\beq
\label{coset:V}
V_{x,\mu} ~=~ U_{x,\mu} \cdot \left[ U^\HP{1}_{x,\mu}\right]^\dagger\,.
\eeq
Apparently it is important here that $U^\HP{1}_{x,\mu}$ is expressed in terms of factually gauge invariant inhomogeneous
coordinates $\omega_x$, $\omega_{x+\mu}$. Note that the results to be discussed below are dependent in principle
upon the particular gauge chosen, at least we don't have precise theoretical arguments to exclude this possibility.
Since we agreed from very beginning not to go deep into theoretical analysis we leave this problem
until the future work~\cite{future}.

In this section we consider the topological aspects of $V$ fields (\ref{coset:V}), investigate their confining
properties and the spectrum of overlap Dirac operator in $V$ fields background.
For this purposes we generated $520$ configurations of $V$ fields from the original data sets at
$\beta=2.400$ and $\beta=2.475$ on $16^4$ lattices.
However, let us start again from the simplest observable $1/2\tr V_p$ and
consider its distribution compared to that of the original fields. The typical graph of this sort,
taken on a particular $\beta=2.475$, $16^4$ configuration,
is presented on Fig.~\ref{fig:coset-plaq}. In fact, for all our $V$ configurations the distributions
of $1/2\tr V_p$ look very similar. Remarkably enough, it is hardly possible to distinguish
the points belonging to the original gauge potentials and to $V$ variables, the only difference
is seen very close to unity and results in $\lesssim 1\%$ larger value of $\langle 1/2\tr V_p\rangle$
on this configuration compared to that of $\langle 1/2\tr U_p\rangle$.  However, it
should not be surprising in view of the above discussed weakness of $\HP{1}$ projected fields.
It might be tempting to conclude that the $V$ fields are essentially the same as the original gauge
potentials and then the validity of $\HP{1}$ projection would become problematic.
However, it is the purpose of this section to show that $V$ fields are completely trivial
as far as the non-perturbative aspects are concerned.

\subsection{Triviality of Topological Properties}
\label{section:coset-topology}

The technically simplest test of topology is provided by the application of $\HP{1}$ $\sigma$-model
embedding method to $V$ configurations.
For us it was almost shocking to discover that the $Q_\HP{1}$ topological charge is identically
zero for all $V$ configurations we have generated on various lattices.
Of course, Eq.~(\ref{coset:V}) expresses our specific
intent to cut away the topology related aspects of the original configurations, however, the identically
vanishing topological charge is indeed remarkable. Evidently this must be checked by other
methods to ensure that $Q \equiv 0$ is not the artifact of $\HP{1}$ approach. We did this
cross-check with overlap-based definition and indeed found that $Q_{overlap}$
is equal to $\pm 1$ only on two $V$ configurations out of $520$.
Evidently, this disagreement of $Q_\HP{1}$ and $Q_{overlap}$ is completely inessential and
should be attributed to what had been discussed in section~\ref{section:numerics}.
Therefore we are confident that the topological properties of $V$ fields are entirely trivial,
the topological charge and susceptibility vanish identically
\beq
Q \equiv 0\,, \qquad \chi \equiv 0\,.
\eeq
This is the first indication that what had been left aside by $\HP{1}$ projection
is the pure perturbation theory. 

\subsection{Loss of Confinement}
\label{section:coset-confinement}

\begin{figure}[t]
\centerline{\psfig{file=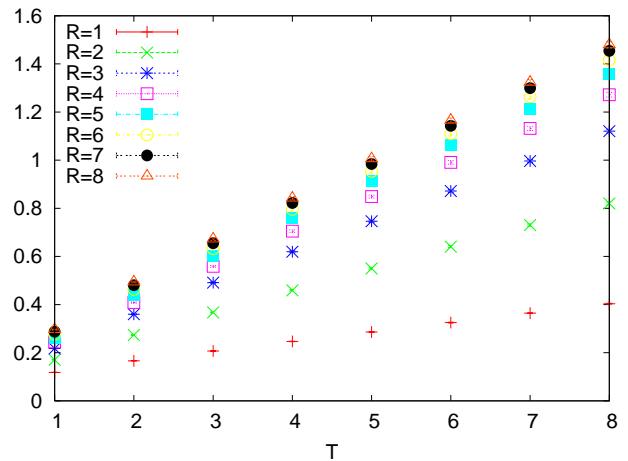,width=0.5\textwidth,silent=,clip=,angle=-90}}
\caption{$-\ln \langle W(T,R) \rangle$ as a function of $T$ at various $R$ measured on $V$
configurations generated on $\beta=2.475$, $16^4$ lattice.}
\label{fig:coset-loops}
\end{figure}

\begin{figure}[t]
\centerline{\psfig{file=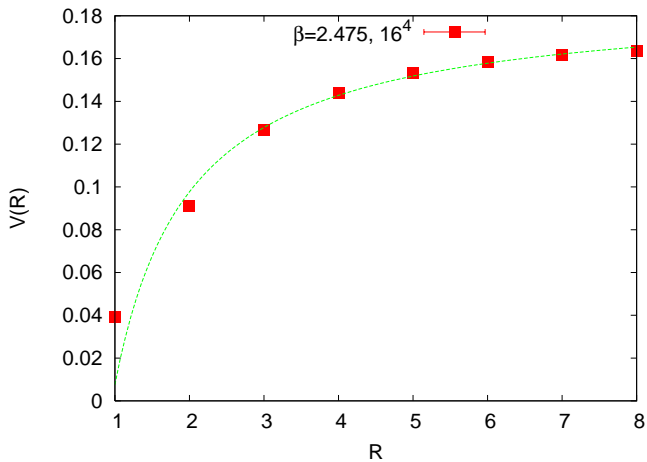,width=0.5\textwidth,silent=,clip=,angle=-90}}
\caption{
The heavy quark potential extracted from $V$ fields on $\beta=2.475$, $16^4$ lattice.
The curve is the fit to analytical dependence (\ref{coset:potential}).}
\label{fig:coset-potential}
\end{figure}

The next question to be addressed is the possibility to have confinement on $V$ configurations
despite of the trivial topology inherent to them. Note that the answer is by no means evident since
the relation between topology and confinement is still not proved rigorously. Consider first the Wilson loops
confinement criterium. We measured the expectation values of planar $T \times R$ Wilson loops
on our $V$ configurations in exactly the same way as was done in section~\ref{section:HP-confinement-wilson}.
As a matter of fact, on all our $V$ data sets the corresponding heavy quark potential
flattens at large distances and  the string tension vanishes. In order to convince the reader that
the potential is indeed constant at large separations we plotted on Fig.~\ref{fig:coset-loops}
the  expectation values $- \ln \langle W(T,R) \rangle$ at various $R$ as a functions of $T$.
It is apparent that the slope of the curves remains constant for $R > 3$ although it is still rising
at smaller distances. The corresponding potential (Fig.~\ref{fig:coset-potential}) was fitted
to analytical equation analogous to (\ref{pot:fit}) and turned out to be totally compatible with Coulomb law
\beq
\label{coset:potential}
V(R) ~=~ const ~+~ \frac{\alpha}{R}\,.
\eeq
However, we do not presenting the best fitted value of $\alpha$ parameter since on our lattices
it would be dictated mostly by $V(R)$ at smallest $R$, which
is highly sensitive to the details of smearing and hypercubic blocking used in our calculations
(note that essentially for this reason the points $R=1,2$ deviate from (\ref{coset:potential})).

As far as the Polyakov line and the corresponding correlation function are concerned, the arguments
of section~\ref{section:HP-confinement-polyakov}  are clearly invalid for $V$ fields
and therefore both $\langle P\rangle$ and $\langle P_0 P_x\rangle$ are expected to be the confinement sensitive observables.
It turns out that the expectation value $\langle P\rangle$ is strictly positive on all our $V$ data sets
being essentially the same at $\beta=2.400$, $\langle P\rangle = 0.294(2)$, and at $\beta=2.475$, $\langle P\rangle = 0.290(4)$.
The strict positivity of Polyakov line distributions apparently rises the question on the relevance
of center symmetry, however, due to the possible lack of statistics we'll not dwell on this issue.
The Polyakov lines connected correlation function $\langle P_0 P_x\rangle - \langle P\rangle^2$
measured on $V$ configurations also clearly shows deconfining behavior, namely, it falls off only as a power of $|x|$
being strictly linear on log-log plot. Note however that the subtraction of the disconnected part violently spoils
the available numerical accuracy and for this reason we do not quote the results of the fits.

To summarize, we found clean evidences that what had been cut away by $\HP{1}$ projection
corresponds to deconfining theory, in which the heavy quark potential is totally compatible
with Coulomb law. This provides the most stringent illustration that the $V$ fields configurations
correspond to pure perturbation theory with no sign whatsoever of neither non-trivial topology
nor confinement. 

\subsection{Fermionic Probes in $V$ Background}
\label{section:coset-fermions}
To conclude this section let us consider the properties of low lying eigenmodes of overlap
Dirac operator in $V$ fields background. The measurements performed here are essentially
the same as was done in section~\ref{section:HP-fermions}. Note that we considered 
only $V$ data sets corresponding to $\beta=2.475$ for reasons to become clear shortly.

\begin{figure}[t]
\centerline{\psfig{file=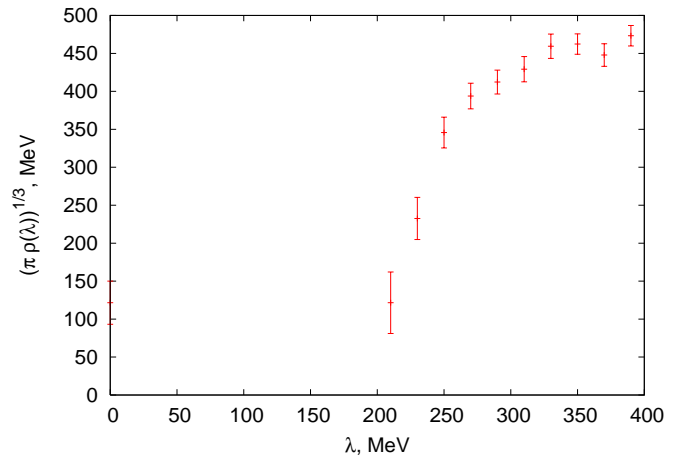,width=0.5\textwidth,silent=,clip=}}
\caption{The spectral density of overlap Dirac operator measured on $V$ fields. Note the complete
absence of eigenmodes in $\lambda \lesssim 200\mathrm{~MeV}$ interval.}
\label{fig:coset-spectrum}
\end{figure}

The spectral density of low lying eigenmodes is presented on Fig.~\ref{fig:coset-spectrum}
and it is apparent that the spectrum is completely empty for $\lambda \lesssim 200 \mathrm{~MeV}$,
no one eigenmode was found in this region on all available $V$ configurations at $\beta=2.475$.
Essentially this precludes any considerations of low lying eigenmodes, there are strictly speaking
no modes to be discussed here. Thus the spectral density for small $\lambda$ is identically zero leading to
vanishing quenched chiral condensate
\beq
\langle\bar\psi\psi\rangle ~=~ 0\,.
\eeq

Evidently it makes no sense to discuss the localization properties of low modes in $V$ background,
the quantities like $I_\lambda$ or $I$, Eq.~(\ref{fermi:IPR}), become undefined.
Nevertheless, it is still instructive to qualitatively consider the change in the degree of
localization for eigenmodes at $\lambda \gtrsim 200 \mathrm{~MeV}$ compared to that on
the original and $\HP{1}$ projected fields (see section~\ref{section:HP-fermions}).
It turns out that the degree of localization expressed by IPRs $I_\lambda$ 
drop down significantly and for all calculated modes $\lambda \gtrsim 200 \mathrm{~MeV}$
it is around $1.2$, which essentially means complete delocalization.

\subsection{Summary}
\label{section:coset-summary}
In this section we presented rather clean and concise evidences that what
had been left aside by $\HP{1}$ projection (\ref{HP:projection}) corresponds to purely perturbation theory.
The primary aim of this presentation was to show that the results of section~\ref{section:HP}
are not invalidated by the components $V_{x,\mu}$ of the original gauge fields which do not survive
$\HP{1}$ projection. We considered this issue from various perspectives and showed that
$V$ fields are not confining, the corresponding heavy quark potential is adequately
described by usual Coulomb law. Moreover, the topological properties of $V$ configurations
are completely trivial, leading, in particular, to the  identically vanishing topological susceptibility.
This result turns out to be quite robust with respect to different definitions of lattice topological charge.
As a consequence the low lying spectrum of overlap Dirac operator in $V$ background
has nothing in common with that on the original or $\HP{1}$ projected fields.
Strictly speaking, the low lying spectrum does not exist, no one eigenmode was found below
$\lambda \lesssim 200 \mathrm{~MeV}$ despite of rather large number of analyzed configurations.
This means that quenched chiral condensate vanishes identically
in $V$ background, $\langle \bar\psi\psi\rangle = 0$.

Therefore, we are confident that the content of the original fields, which is cut away by $\HP{1}$
projection, is completely trivial as far as the non-perturbative aspects are concerned. This provides
at least numerically solid grounds for the results obtained in section~\ref{section:HP}.

\section{Conclusions}
\label{section:conclusions}
The primary goal of the present publication was the further development of $\HP{1}$ $\sigma$-model
embedding approach~\cite{Gubarev:2005rs} aimed to investigate the topology of $SU(2)$ gauge fields.
While Ref.~\cite{Gubarev:2005rs} discussed in details the theoretical aspects, the tests of actual
numerical implementation were not so much convincing. In this paper we filled this gap and presented
the high statistics comparison of $\HP{1}$ based topological charge, $Q_\HP{1}$, with ones obtained via
field-theoretical, $Q_{clover}$, and overlap Dirac operator based, $Q_{overlap}$, approaches.
As far as the field-theoretical definition is concerned, the corresponding comparison is meaningful
only on semiclassical configurations and in this case we essentially found the complete equivalence
of $Q_\HP{1}$ and $Q_{clover}$. In the most interesting case of equilibrium vacuum configurations
we confronted $Q_\HP{1}$ and $Q_{overlap}$ on rather large number of data sets at different spacings
and found that both definitions are strongly correlated and identify essentially the same topology.
Unfortunately, it was difficult to compare our results with similar calculations in the literature,
however, the correlation between $Q_\HP{1}$ and $Q_{overlap}$ turns out to be quantitatively the same
as it was in the case of $Q_{overlap}$ and $Q_{clover}$ definitions. Basing on that we concluded
that the $\HP{1}$ $\sigma$-model embedding method is, in fact, computationally superior to any other
definitions of the lattice topological charge available so far.

The next immediate problem is to investigate on the numerically confident level the scaling
of the topological susceptibility both with lattice spacing and volume. Besides of the methodological
importance, this problem is crucial in order to address the physical relevance of the method
itself and its sensitivity to lattice artifacts.  We performed statistically
convincing scaling checks of the topological susceptibility using the $\HP{1}$ embedding approach
and found almost perfect scaling with both lattice spacing and volume.
The topological susceptibility being extrapolated to the continuum limit
turns out to be $\chi^{1/4} = 216(4)\mathrm{~MeV}$ and is totally compatible with what is known in the literature.
The scaling properties of $\chi$ make us confident that the $\HP{1}$ embedding
method is insensitive to lattice dislocations, moreover, the topological susceptibility
does not require, in fact, neither multiplicative nor additive renormalizations.

As might be anticipated from the results of Ref.~\cite{Gubarev:2005rs}, the $\HP{1}$ embedding
approach is not restricted to the topology related measurements. Its essence is to provide
a unique (modulo Gribov copies problem which, however, was proved to be inessential)
configuration of quaternionic valued scalar fields to any given original gauge background.
Then the corresponding $\HP{1}$ projection comes out naturally and we considered in details
the properties of $\HP{1}$ projected fields. Note that $\HP{1}$ projection is radically
different from what is commonly known by this term, in particular because it maintains the
gauge invariance from very beginning. The first striking observation here is that
the mean curvature of projected potentials is extremely small, while its spacing
dependence shows no sign whatsoever of the perturbation theory. In fact, it only contains
the terms which are to be identified with gluon condensate and non-trivial quadratic correction
to $\langle\alpha_s\,G^2/\pi\rangle$. We calculated both this quantities with inaccessible
so far accuracy, the result for the gluon condensate being 
$\langle\alpha_s\,G^2/\pi\rangle = 0.066(2) \mathrm{~GeV}^4$
which fairly agrees with the existing literature. On the other hand, the numerically
precise value of the quadratic correction satisfies the known tight bounds and  allows to pose the
problem of unusual power corrections in Yang-Mills theory on qualitatively new level.

Then we addressed the crucial problem of confinement in the language of $\HP{1}$ projection.
It is remarkable that despite of the weakness of the projected fields the $\HP{1}$ string tension is not
only non-zero at finite lattice spacing, it accounts, in fact, for full $SU(2)$ string tension in the
continuum limit, $\lim_{a\to 0} [\sigma^\HP{1}/\sigma^{SU(2)}]^{1/2} = 1.04(3)$.
Moreover, the explicitness of the gauge fields topology in $\HP{1}$ variables
allows us to study the microscopic origin of the projected string tension.
We argued that $\sigma^\HP{1} \ne 0$  is to be associated with the existence of percolating
regions of sign-coherent topological charge, first discovered (with overlap-based approach) in~\cite{Horvath-structures}
and then confirmed in $\HP{1}$ method in~\cite{Gubarev:2005rs}. As far as we can see, this is the first
explicit demonstration of the confinement-topology interrelation.

We concluded the investigation of $\HP{1}$ projected fields with consideration of low lying spectrum of
overlap Dirac operator in the projected background. The spectral density of low eigenmodes turns
out to be essentially the same as it was on the original gauge fields and allows us to estimate
the value of chiral condensate in the quenched approximation $\langle\bar\psi\psi\rangle^{1/3} = 278(6)\mathrm{~MeV}$,
which is fairly compatible with its commonly accepted value. Simultaneously we measured
the localization properties of low eigenmodes, which are conventionally expressed in term of
the inverse participation ratio (IPR). Knowledge of IPRs at various spacings allows us to conclude
that the localization regions are likely to be either two- or three-dimensional.

It was understood from very beginning that the validity of the above results relies strongly
on the ability of $\HP{1}$ projection to capture {\it only} non-perturbative aspects of the
original gauge fields. Therefore it is mandatory to study what had been left aside by $\HP{1}$ projection.
We investigated carefully this issue by showing first that the part of the gauge fields not surviving
$\HP{1}$ projection could be defined unambiguously. Then it is the matter of straightforward calculation
to show that indeed what had been cut away by $\HP{1}$ projection is completely trivial as far
as the non-perturbative aspects are concerned. In particular, there is no topology in these fields,
the topological charge and susceptibility vanish identically; confinement is definitely
lost although the Coulomb part of the potential is clearly seen; and finally the low Dirac spectrum
completely disappears, not a single low eigenmode was found on our rather large set of configurations.
Thus we concluded that the $\HP{1}$ projection indeed captures only the non-perturbative content of gauge
background and what is left aside corresponds to pure perturbation theory.

Finally let us remind our original intent not to discuss the theoretical implications
of the above findings. Only experimental numerical results had been reported since otherwise the discussions
would definitely lead us far beyond the scope of present publication. The corresponding theoretical treatment
is to be published elsewhere~\cite{future}.

\section*{Acknowledgments}
The authors are grateful to the members of ITEP lattice group for stimulating
discussions. The work was partially
supported  by grants RFBR-05-02-16306a, RFBR-05-02-17642, RFBR-0402-16079 and
RFBR-03-02-16941. F.V.G. was partially supported by INTAS YS grant 04-83-3943.



\begin{thebibliography}{99}

\bibitem{Neuberger:1997fp}
  H.~Neuberger, Phys.\ Lett.\ B {\bf 417}, 141 (1998); Phys.\ Lett.\ B {\bf 427}, 353 (1998).

\bibitem{Atiyah:1971rm}
  M.~F.~Atiyah and I.~M.~Singer, Annals Math.\  {\bf 93} (1971) 139.

\bibitem{overlap}
  R.~Narayanan and H.~Neuberger, Nucl.\ Phys.\ B {\bf 443}, 305 (1995);\\
  P.~Hasenfratz, V.~Laliena and F.~Niedermayer, Phys.\ Lett.\ B {\bf 427}, 125 (1998);\\
  D.~H.~Adams, Annals Phys.\  {\bf 296}, 131 (2002);\\
  R.~Narayanan and P.~M.~Vranas, Nucl.\ Phys.\ B {\bf 506}, 373 (1997).

\bibitem{Horvath-structures}
  I.~Horvath {\it et al.}, Phys.\ Lett.\ B {\bf 612}, 21 (2005);
                           Phys.\ Rev.\ D {\bf 68}, 114505 (2003);
                           Nucl.\ Phys.\ Proc.\ Suppl.\  {\bf 129}, 677 (2004);
  A.~Alexandru, I.~Horvath and J.~b.~Zhang, Phys.\ Rev.\ D {\bf 72}, 034506 (2005).

\bibitem{Gubarev:2005rs}
  F.~V.~Gubarev and S.~M.~Morozov, Phys.\ Rev.\ D {\bf 72}, 76008 (2005).

\bibitem{Atiyah:1978ri}
  M.~F.~Atiyah, N.~J.~Hitchin, V.~G.~Drinfeld and Y.~I.~Manin,  Phys.\ Lett.\ A {\bf 65}, 185 (1978).

\bibitem{Drinfeld:1978xr}
  V.~G.~Drinfeld and Y.~I.~Manin, Commun.\ Math.\ Phys.\  {\bf 63}, 177 (1978).

\bibitem{HPN-various}
  F.~Gursey and C.~H.~Tze, Annals Phys.\  {\bf 128}, 29 (1980).
  J.~Lukierski,  CERN-TH-2678; \\
  Y.~N.~Kafiev, Phys.\ Lett.\ B {\bf 87} (1979) 219; Phys.\ Lett.\ B {\bf 96} (1980) 337;
                Nucl.\ Phys.\ B {\bf 178}, 177 (1981);\\
  M.~A.~Jafarizadeh, M.~Snyder and C.~H.~Tze, Nucl.\ Phys.\ B {\bf 176}, 221 (1980);\\
  D.~Maison, MPI-PAE/PTh 52/79;\\
  J.~E.~Avron, L.~Sadun, J.~Segert and B.~Simon,  Comm.\ Math.\ Phys.\  {\bf 124}, 595 (1989);\\
  M.~T.~Johnsson and I.~J.~R.~Aitchison,  J.\ Phys.\ A {\bf 30}, 2085 (1997);\\
  E.~Demler and S.~C.~Zhang,  Annals Phys.\  {\bf 271}, 83 (1999).

\bibitem{Dubrovin}
  B.~A.~Dubrovin, A.~T.~Fomenko, S.~P.~Novikov,
  {\it ``Modern Geometry: Methods and Applications''}, 
  {\it ``Modern Geometry: Introduction to Homology Theory''},
  New York: Springer-Verlag (1992).

\bibitem{Gubarev:2005jm}
  F.~V.~Gubarev, S.~M.~Morozov, M.~I.~Polikarpov and V.~I.~Zakharov,
  {\it ``Low lying eigenmodes localization for chirally symmetric Dirac operator''}, hep-lat/0505016,
  to be published in JETP Letters.

\bibitem{Polikarpov:2005ey}
  M.~I.~Polikarpov, F.~V.~Gubarev, S.~M.~Morozov and V.~I.~Zakharov,
           {\it 'Localization of low lying eigenmodes for chirally symmetric Dirac operator'},
           arXiv:hep-lat/0510098.

\bibitem{A2}
  F.~V.~Gubarev, L.~Stodolsky and V.~I.~Zakharov, Phys.\ Rev.\ Lett.\  {\bf 86}, 2220 (2001);\\
  F.~V.~Gubarev and V.~I.~Zakharov, Phys.\ Lett.\ B {\bf 501}, 28 (2001);\\
  D.~Dudal, H.~Verschelde, J.~A.~Gracey, V.~E.~R.~Lemes, M.~S.~Sarandy, R.~F.~Sobreiro and S.~P.~Sorella,
                                 JHEP {\bf 0401}, 044 (2004);\\
  F.~V.~Gubarev and S.~M.~Morozov, Phys.\ Rev.\ D {\bf 71}, 114514 (2005);\\
  D.~V.~Bykov and A.~A.~Slavnov,
        {\it '``Dimension two vacuum condensates in gauge-invariant theories'}, arXiv:hep-th/0505089;\\
  X.~d.~Li and C.~M.~Shakin, Phys.\ Rev.\ D {\bf 71}, 074007 (2005);\\
  D.~Dudal, R.~F.~Sobreiro, S.~P.~Sorella and H.~Verschelde, Phys.\ Rev.\ D {\bf 72}, 014016 (2005).

\bibitem{A2-review}
  V.~I.~Zakharov, AIP Conf.\ Proc.\  {\bf 756}, 182 (2005);\\
  V.~I.~Zakharov, Phys.\ Usp.\  {\bf 47} (2004) 37.

\bibitem{viz}
  F.~V.~Gubarev, M.~I.~Polikarpov and V.~I.~Zakharov, Surveys High Energ.\ Phys.\  {\bf 15}, 89 (2000);\\
  V.~I.~Zakharov, {\it 'From confining fields back to power corrections'}, arXiv:hep-ph/0509114;\\
  V.~I.~Zakharov, {\it 'Hints on dual variables from the lattice SU(2) gluodynamics'}, arXiv:hep-ph/0309301.

\bibitem{viz2}
  V.~I.~Zakharov, {\it 'Confining fields in lattice SU(2)'}, arXiv:hep-ph/0312210;\\
  V.~I.~Zakharov, {\it 'Non-perturbative match of ultraviolet renormalon'}, arXiv:hep-ph/0309178.

\bibitem{future}
  P.Yu.~Boyko, F.V.~Gubarev, S.M.~Morozov,  in preparation.

\bibitem{vortex-removal}
  P.~de Forcrand and M.~D'Elia, {\it Phys. Rev. Lett.} {  \bf 82}, 4582 (1999);\\
  J. Gattnar et al., {\it Nucl.Phys.} {\bf B716}, 105 (2005);\\
  A.~V.~Kovalenko, M.~I.~Polikarpov, S.~N.~Syritsyn and V.~I.~Zakharov, Phys.\ Lett.\ B {\bf 613}, 52 (2005).

\bibitem{Bornyakov:2005wp}
  V.~G.~Bornyakov, M.~I.~Polikarpov, T.~Suzuki and S.~N.~Syritsyn, AIP Conf.\ Proc.\  {\bf 756} (2005) 463.

\bibitem{Kronfeld:1986ts}
  A.~S.~Kronfeld, M.~L.~Laursen, G.~Schierholz and U.~J.~Wiese,  Nucl.\ Phys.\ B {\bf 292}, 330 (1987).

\bibitem{Lucini:2001ej}
  B.~Lucini and M.~Teper, JHEP {\bf 0106}, 050 (2001).

\bibitem{spacings}
  J.~Fingberg, U.~M.~Heller and F.~Karsch, Nucl.\ Phys.\ B {\bf 392}, 493 (1993);\\
  G.~S.~Bali, K.~Schilling and A.~Wachter, Phys.\ Rev.\ D {\bf 55}, 5309 (1997);\\
  G.~S.~Bali, K.~Schilling and C.~Schlichter, Phys.\ Rev.\ D {\bf 51}, 5165 (1995).

\bibitem{DiVecchia:1981qi}
  P.~Di Vecchia, K.~Fabricius, G.~C.~Rossi and G.~Veneziano,  Nucl.\ Phys.\ B {\bf 192}, 392 (1981);
  Phys.\ Lett.\ B {\bf 108}, 323 (1982).

\bibitem{Cundy:2002hv}
  N.~Cundy, M.~Teper and U.~Wenger, Phys.\ Rev.\ D {\bf 66}, 094505 (2002).

\bibitem{GarciaPerez:1998ru}
  M.~Garcia Perez, O.~Philipsen and I.~O.~Stamatescu, Nucl.\ Phys.\ B {\bf 551}, 293 (1999).

\bibitem{Bilson-Thompson:2002jk}
  S.~O.~Bilson-Thompson, D.~B.~Leinweber and A.~G.~Williams, Annals Phys.\  {\bf 304}, 1 (2003).

\bibitem{deForcrand:1997sq}
  P.~de Forcrand, M.~Garcia Perez and I.~O.~Stamatescu, Nucl.\ Phys.\ B {\bf 499}, 409 (1997).

\bibitem{Giusti:2002sm}
  L.~Giusti, C.~Hoelbling, M.~Luscher and H.~Wittig, Comput.\ Phys.\ Commun.\  {\bf 153}, 31 (2003).

\bibitem{zenkin}
  T.~W.~Chiu and S.~V.~Zenkin, Phys.\ Rev.\ D {\bf 59}, 074501 (1999);\\
  T.~W.~Chiu, C.~W.~Wang and S.~V.~Zenkin, Phys.\ Lett.\ B {\bf 438}, 321 (1998).

\bibitem{Capitani:1999uz}
  S.~Capitani, M.~Gockeler, R.~Horsley, P.~E.~L.~Rakow and G.~Schierholz, Phys.\ Lett.\ B {\bf 468}, 150 (1999).

\bibitem{Giusti:2003gf}
  L.~Giusti, M.~Luscher, P.~Weisz and H.~Wittig, JHEP {\bf 0311}, 023 (2003).

\bibitem{Shifman:1978bx}
  M.~A.~Shifman, A.~I.~Vainshtein and V.~I.~Zakharov, Nucl.\ Phys.\ B {\bf 147}, 385 (1979);
                                                      Nucl.\ Phys.\ B {\bf 147}, 448 (1979).
\bibitem{Rakow:2005yn}
  P.~E.~L.~Rakow, {\it 'Stochastic perturbation theory and the gluon condensate'}, arXiv:hep-lat/0510046.

\bibitem{Baig:1985pu}
  M.~Baig, {\it 'Gluon Condensation Parameter In SU(2) Lattice Gauge Theory And Large N Universality'}, UAB-FT-124.

\bibitem{DiGiacomo:1981wt} 
  A.~Di Giacomo and G.~C.~Rossi, Phys.\ Lett.\ B {\bf 100}, 481 (1981).

\bibitem{gluon}
  T.~Banks, R.~Horsley, H.~R.~Rubinstein and U.~Wolff, Nucl.\ Phys.\ B {\bf 190}, 692 (1981);\\
  S.~s.~Xue, Phys.\ Lett.\ B {\bf 191}, 147 (1987);\\
  B.~Alles, M.~Campostrini, A.~Feo and H.~Panagopoulos, Phys.\ Lett.\ B {\bf 324}, 433 (1994);\\
  A.~Di Giacomo,{\it 'Non perturbative QCD'}, arXiv:hep-lat/9912016;\\
  K.~Langfeld, E.~M.~Ilgenfritz and H.~Reinhardt, Nucl.\ Phys.\ B {\bf 608}, 125 (2001).

\bibitem{A2-old}
  J.~Greensite and M.~B.~Halpern, Nucl.\ Phys.\ B {\bf 271}, 379 (1986);\\
  M.~J.~Lavelle and M.~Schaden, Phys.\ Lett.\ B {\bf 208}, 297 (1988);\\
  M.~Lavelle and M.~Oleszczuk, Mod.\ Phys.\ Lett.\ A {\bf 7} (1992) 3617.

\bibitem{Burgio:1997hc}
  G.~Burgio, F.~Di Renzo, G.~Marchesini and E.~Onofri, Phys.\ Lett.\ B {\bf 422}, 219 (1998).

\bibitem{A2-plaq}
  F.~Di Renzo and L.~Scorzato, JHEP {\bf 0110}, 038 (2001);\\
  R.~Horsley, P.~E.~L.~Rakow and G.~Schierholz, Nucl.\ Phys.\ Proc.\ Suppl.\  {\bf 106}, 870 (2002).

\bibitem{Albanese:1987ds}
  M.~Albanese {\it et al.}  [APE Collaboration], Phys.\ Lett.\ B {\bf 192}, 163 (1987).

\bibitem{Hasenfratz:2001hp}
  A.~Hasenfratz and F.~Knechtli, Phys.\ Rev.\ D {\bf 64}, 034504 (2001) [arXiv:hep-lat/0103029].

\bibitem{Bali:1994de}
  G.~S.~Bali, K.~Schilling and C.~Schlichter, Phys.\ Rev.\ D {\bf 51}, 5165 (1995).

\bibitem{A2-confine}
  K.~Ishiguro {\it et al.},
     {\it 'Towards SU(2) invariant formulation of the monopole confinement mechanism'}, arXiv:hep-lat/0509089;\\
  V.~I.~Shevchenko, {\it 'A remark on the short distance potential in gluodynamics'}, arXiv:hep-ph/0301280;\\
  K.~I.~Kondo, Phys.\ Lett.\ B {\bf 514}, 335 (2001);\\
  H.~Verschelde, K.~Knecht, K.~Van Acoleyen and M.~Vanderkelen, Phys.\ Lett.\ B {\bf 516}, 307 (2001);\\
  T.~Suzuki, K.~Ishiguro, Y.~Mori and T.~Sekido, Phys.\ Rev.\ Lett.\  {\bf 94}, 132001 (2005);\\
  M.~N.~Chernodub {\it et al.}, Phys.\ Rev.\ D {\bf 72}, 074505 (2005).


\bibitem{Gattringer}
  C.~Gattringer, M.~Gockeler, P.~E.~L.~Rakow, S.~Schaefer and A.~Schafer,
                           Nucl.\ Phys.\ B {\bf 617}, 101 (2001);
                           Nucl.\ Phys.\ B {\bf 618}, 205 (2001).

\bibitem{DeGrand}
  T.~DeGrand and A.~Hasenfratz, Phys.\ Rev.\ D {\bf 64}, 034512 (2001);
                                Phys.\ Rev.\ D {\bf 65}, 014503 (2002).

\bibitem{Hip}
  I.~Hip, T.~Lippert, H.~Neff, K.~Schilling and W.~Schroers, Phys.\ Rev.\ D {\bf 65}, 014506 (2002);
  R.~G.~Edwards and U.~M.~Heller, Phys.\ Rev.\ D {\bf 65}, 014505 (2002);
  T.~Blum {\it et al.}, Phys.\ Rev.\ D {\bf 65}, 014504 (2002).

\bibitem{Horvath}
  I.~Horvath {\it et al.}, Phys.\ Rev.\ D {\bf 66}, 034501 (2002);\\
  I.~Horvath, N.~Isgur, J.~McCune and H.~B.~Thacker, Phys.\ Rev.\ D {\bf 65}, 014502 (2002).

\bibitem{McLerran:1981pb}
  L.~D.~McLerran and B.~Svetitsky, Phys.\ Rev.\ D {\bf 24}, 450 (1981).

\bibitem{Gross:1974jv}
  D.~J.~Gross and A.~Neveu, Phys.\ Rev.\ D {\bf 10}, 3235 (1974).

\bibitem{fermions-confinement}
  E.~Witten, Nucl.\ Phys.\ B {\bf 149}, 285 (1979); Nucl.\ Phys.\ B {\bf 156}, 269 (1979);\\
  A.~Casher, Phys.\ Lett.\ B {\bf 83}, 395 (1979).

\bibitem{fermions-reviews}
  H.~G.~Dosch and Y.~A.~Simonov, Phys.\ Atom.\ Nucl.\  {\bf 57}, 143 (1994) [Yad.\ Fiz.\  {\bf 57N1}, 152 (1994)];\\
  Y.~A.~Simonov,  {\it 'Chiral symmetry breaking and confinement in QCD'},
                  talk given at 31st International Ahrenshoop Symposium on the Theory of Elementary Particles,
                  Buckow, Germany, 2-6 Sep 1997;\\
  K.~E.~Cahill,   {\it 'Chiral symmetry and quark confinement'}, arXiv:hep-ph/9812312.

\bibitem{banks-casher}
  T. Banks, A. Casher, Nucl.\ Phys.\ B {\bf 169}, 103 (1980).

\bibitem{IPR-defs}
  C.~Aubin {\it et al.}  [MILC Collaboration],
           {\it 'The scaling dimension of low lying Dirac eigenmodes and of the  topological charge density'},
           arXiv:hep-lat/0410024;\\
  C.~Gattringer, M.~Gockeler, P.~E.~L.~Rakow, S.~Schaefer and A.~Schafer, Nucl.\ Phys.\ B {\bf 617}, 101 (2001);\\
  Y.~Koma, E.~M.~Ilgenfritz, K.~Koller, G.~Schierholz, T.~Streuer and V.~Weinberg,
           Proc.\ Sci.\  {\bf LAT2005}, 300 (2005);\\
  C.~Bernard {\it et al.}, {\it 'More evidence of localization in the low-lying Dirac spectrum'},
           arXiv:hep-lat/0510025;\\


\bibitem{Greensite:2005dv}
  J.~Greensite, S.~Olejnik, M.~Polikarpov, S.~Syritsyn and V.~Zakharov, Phys.\ Rev.\ D {\bf 71}, 114507 (2005);
                               Proc.\ Sci.\  {\bf LAT2005}, 325 (2005), [arXiv:hep-lat/0509070].

\bibitem{psi-psi}
  S.~J.~Hands and M.~Teper, Nucl.\ Phys.\ B {\bf 347}, 819 (1990);\\
  E.~Follana, A.~Hart, C.~T.~H.~Davies and Q.~Mason  [HPQCD Collaboration], Phys.\ Rev.\ D {\bf 72}, 054501 (2005);\\
  L.~Giusti, F.~Rapuano, M.~Talevi and A.~Vladikas, Nucl.\ Phys.\ B {\bf 538}, 249 (1999).
\end{thebibliography}
\end{document}